\documentclass{article}

\usepackage{arxiv}

\usepackage[utf8]{inputenc} 
\usepackage[T1]{fontenc}    
\usepackage{hyperref}       
\usepackage{url}            
\usepackage{booktabs}       
\usepackage{amsfonts}       
\usepackage{nicefrac}       
\usepackage{microtype}      
\usepackage{graphicx}

\title{Beyond Driving: Envisioning Activities in Future Autonomous Vehicles through Experience-Centered Design}

\date{September 2026}

\hypersetup{
  pdftitle={Beyond Driving: Envisioning Activities in Future Autonomous Vehicles through Experience-Centered Design},
  pdfauthor={Keqi Chen, Xiao Xue, Xinyi Liu, Runjia Tan, Chris Speed, Lee Kwan Min, and Chen Lv},
  pdfkeywords={Autonomous Vehicles; Non-Driving-Related Activities; Experience-Centered Design; Travel Time Use; Activity-Based Travel},
  hidelinks
}

\author{
 Keqi Chen$^{1}$, Xiao Xue$^{2}$, Xinyi Liu$^{3}$, Runjia Tan$^{1}$,\\
 Chris Speed$^{4}$, Lee Kwan Min$^{5}$, and Chen Lv$^{1,*}$\\[0.55em]
 \normalfont\small
 $^{1}$School of Mechanical and Aerospace Engineering, Nanyang Technological University, Singapore\\
 $^{2}$The Future Laboratory, Tsinghua University, Beijing, China\\
 $^{3}$Singapore American School, Singapore\\
 $^{4}$Design for Regenerative Futures, RMIT, Melbourne, Australia\\
 $^{5}$Wee Kim Wee School of Communication and Information, Nanyang Technological University, Singapore\\[0.35em]
 $^{*}$Corresponding author: \texttt{lyuchen@ntu.edu.sg}
}

\begin{document}
\maketitle
\begin{abstract}
Autonomous vehicles (AVs) are poised to fundamentally alter personal transportation, offering occupants the freedom to engage in various non-driving-related activities (NDRAs). However, our current understanding of how people might actually use this time in fully autonomous vehicles (FAVs) is limited. Traditional research methods often struggle to capture the influence of diverse travel circumstances and purposes when exploring future scenarios. This paper introduces experience-centered design (ECD) as an approach to investigate potential NDRAs within FAVs by examining the intricate connections between individuals' daily routines, specific travel contexts, and the activities they might undertake in transit. Through a multi-phase study employing participatory techniques that facilitated narrative construction and exploration—including diary studies, scenario scripting, and mixed reality (MR) enactments—we enabled participants to ground speculative future scenarios in their own lived experiences. This process yielded nuanced insights into preferences and behaviors concerning potential NDRAs, alongside the underlying subjective meanings and sociotechnical considerations. Our findings lead us to conceptualize NDRAs not as isolated instances of "travel time use," but as dynamic sequences of interrelated activities deeply shaped by pre- and post-journey contexts. The effectiveness of our ECD approach in bridging current lived experiences with future scenarios was crucial for uncovering these insights. Ultimately, this study reconceptualizes AVs as complex sociotechnical systems that actively mediate human activity and interaction, suggesting a fundamental shift in their role within the urban fabric.
\end{abstract}

\keywords{Autonomous Vehicles \and Non-Driving-Related Activities \and Experience-Centered Design \and Travel Time Use \and Activity-Based Travel}

\section{Introduction}

The growing adoption of Autonomous Vehicles (AVs)~\cite{sae2014international, lipson2016driverless, litman2017autonomous} is poised to fundamentally change daily commutes by redefining vehicle occupants' roles and experiences. As automation advances, traditional drivers undergo what Mokhtarian (2018) describes as 'passengerisation'~\cite{mokhtarian2018times}. This shift means time previously devoted to driving can be reallocated~\cite{gripsrud2012working}, potentially transforming transit periods into opportunities for diverse engagement. Consequently, researchers have begun exploring how people might engage in non-driving-related activities (NDRAs) in fully autonomous vehicles (FAVs), leading to various approaches for envisioning future in-vehicle scenarios.

Despite numerous studies acknowledging the importance of anticipating activities within future AV travel~\cite{schoettle2014survey,kyriakidis2015public,bansal2016assessing,wadud2019fully}, a significant gap remains in understanding the dynamic relationship between specific NDRAs and the highly variable contexts of travel. These contexts include trip purpose, time constraints, social settings, the sequence of activities before and after journeys, and available in-vehicle resources, all varying significantly across different traveler profiles and needs. While concepts like 'travel time utilization' and 'travel-based multitasking'~\cite{konig2017users} are common, particularly in Information and Communication Technologies (ICT) research, critics highlight their limitations. These perspectives often focus on generalized opinions about 'undirected travel'~\cite{mokhtarian2001derived}, conceptually separating travel time from daily life and neglecting crucial spatiotemporal relationships between travel segments and broader activity contexts~\cite{chu2012review}. As AVs free individuals from driving tasks, travel time likely becomes more integrated with, rather than separate from, an individual's broader routine-based considerations and schedule demands~\cite{lyons2005travel}, reflecting how travel demand derives from broader life activities~\cite{bhat2000multi}. However, determining which contextual factors most significantly influence NDRA choices and understanding how these factors interact within integrated daily routines presents a significant challenge, largely because realistically situating participants within these inherently uncertain future scenarios proves difficult with conventional methods.

Existing studies rely heavily on "stated intention" methods~\cite{schoettle2014survey,kyriakidis2015public,bansal2016assessing,wadud2019fully}, for example, primarily using focus groups and questionnaires, which risk collecting speculative responses, as concrete daily AV experiences remain unfamiliar to most people. However, these methods can face challenges in effectively linking participants' current travel experiences and routines to potential NDRA engagement within personally relevant future AV scenarios. As some researchers caution, relying solely on stated intentions for unfamiliar technologies might not adequately create platforms where participants can draw upon their lived experiences, thoughts, feelings, and values through both discussion and action~\cite{sanders2008co, sheeran2002intention}, potentially leading to disparities between anticipated and actual future behaviors~\cite{sorensen1992thought, yang2020chicken}. While some notable work has made progress, such as Stevens et al. (2019), who used participatory design workshops to co-create future AV interior concepts based on user activities~\cite{stevens2019using}, and Hecht et al. (2020), who employed technology-augmented driving simulators to evoke affective responses to potential AV situations~\cite{hecht2020you}, these investigations often presented somewhat generalized travel scenarios that did not fully embed the simulated journey within participants' complex, unique daily activity patterns. As Human-Computer Interaction (HCI) research on technology futuring emphasizes creating "discursive spaces"~\cite{lindley2015back} for participation and critique, we need a methodological approach focusing on participatory and context-sensitive exploration. This approach should enable participants to actively consider and "rehearse" future possibilities grounded in their own lives, leading to a more comprehensive understanding of future AV circumstances.

This study reports methodological and conceptual insights from innovative workshops coupled with an AV cabin and mixed reality (MR) scenarios. We adopted experience-centered design (ECD) and implemented it through a three-phase methodology consisting of a one-week pre-workshop sensitization, scriptwriting, and scene enactment to capture connections between individuals' travel experiences and NDRAs in future FAVs. This methodology catalyzes participatory inquiry by having participants share and reflect on their travel experiences through real enactments, illuminating contextual factors that may influence NDRAs beyond a single trip. Our study makes several key contributions to future-oriented AV research:
\begin{enumerate}
\item We demonstrate how ECD effectively bridges current experiences and future possibilities beyond traditional methods. Through tools like the Journey Canvas and a systematic process of sensitization, storytelling, and embodied enactment, we showcase patterns that enable participants to ground future scenarios in lived experiences while revealing rich insights about potential AV interactions.
\item We reconceptualize NDRAs in autonomous vehicles as dynamic sequences of activities shaped by broader temporal and spatial contexts. This challenges conventional "travel time use" paradigms and suggests new ways of understanding how vehicle spaces might be experienced and utilized in autonomous futures.
\item We illuminate how potential users perceive and negotiate AVs as complex sociotechnical systems within their lived contexts. This perspective reveals AVs not merely as transport tools but as emergent spaces that could mediate experiences and interactions within the city, highlighting the crucial interplay between technology, context, and human values in shaping future mobility.
\end{enumerate}

\section{Related Work} 
\subsection{Paradigm Shift: From "Travel-based Activity" to "Activity-based Travel"}
The concept of "Travel-based activity" originated within transportation planning models, specifically in the "Trip-based travel demand model" developed by the Metropolitan Transportation Commission (MTC)~\cite{ruiter1978disaggregate}. This approach frames travel as discrete trips between locations, focusing on the journey itself rather than the activities motivating it. Within this framework, activities during travel (such as reading or working) are considered secondary to the primary purpose of moving between points. This perspective may have potentially shaped the framing of numerous studies on NDRAs in AV research~\cite{schoettle2014survey,kyriakidis2015public, bansal2016assessing, cunningham2019public}, where researchers primarily examine which activities might occur during travel without necessarily connecting them to broader life patterns. Since the 1970s, however, a significant paradigm shift has occurred toward "Activity-Based Travel" models. This shift is evident in several landmark studies: Chapin's work (1974) on human activity patterns as the foundation for travel demand~\cite{chapin1974human}, Bowman and Ben-Akiva's (2001) development of activity-based discrete choice models~\cite{bowman2001activity}, Kitamura's (1988) evaluation of activity-based travel analysis methodologies~\cite{kitamura1988evaluation}, Bhat and Koppelman's (1999) comprehensive review of activity-based modeling approaches~\cite{bhat1999activity}, and McNally's (2000) framework integrating activity patterns with travel behavior analysis~\cite{mcnally2000activity}. Together, these studies demonstrate how the field has evolved toward understanding travel as derived from and embedded within daily activity patterns. 

The Activity-Based Travel Demand Model, as formulated by Axhausen and Garling (1992), establishes travel as a derived demand—not an end in itself but a means to participate in activities distributed across space~\cite{axhausen1992activity}. Hagerstrand's time-geography theory (1970) enriches this perspective by analyzing how personal, coupling (interpersonal), and authority (institutional) constraints shape individuals' spatiotemporal activity patterns~\cite{ilagcrstrand1970people}. From this theoretical foundation, travel decisions emerge from complex optimization problems where individuals fulfill activity needs while navigating constraints including time budgets, financial resources, and social obligations. Furthermore, insights from behavioral economics and psychology highlight that travel behavior isn't purely instrumental. Studies reveal the intrinsic positive utility some find in travel~\cite{mokhtarian2001derived}, the influence of psychological factors like autonomy and status~\cite{metz2008myth}, and the reflection of broader personal preferences in travel-time activities~\cite{ettema2007multitasking}. Connections between travel satisfaction, well-being, and activity participation further reveal travel's affective dimensions~\cite{de2013travel}, showing how it intertwines with users' emotional states and overall life experience.

In the context of AVs, the traditional boundaries between "travel time" and "activity time" increasingly blur~\cite{lyons2005travel}, with research by Singleton (2019) suggesting that AVs may fundamentally alter how users perceive and value travel time~\cite{singleton2019discussing}. Burns et al. (2013) and Krueger et al. (2016) have documented emerging behavioral patterns where potential AV users anticipate reorganizing their activity schedules around new travel capabilities~\cite{burns2013transforming, krueger2016preferences}. This transformation highlights the necessity of studying AV travel within a comprehensive spatiotemporal continuum that encompasses users' entire daily activity patterns rather than isolating it as distinct "travel time." Despite this established theoretical understanding, practical explorations and design-led inquiries within HCI that examine how potential NDRAs within AVs integrate with users’ actual activity patterns remain relatively scarce. This gap suggests a need not only for methodological innovation but perhaps also for revisiting core paradigms; indeed, understanding future AV use in everyday life may demand perspectives that extend beyond established Activity-Based models, a possibility explored through the approach presented in this paper.

\subsection{NDRAs in AVs: Contextual Insights and Methods}
Research on NDRAs in AVs has expanded considerably in recent years, with numerous studies exploring how passengers might spend their time during autonomous journeys~\cite{schoettle2014survey, kyriakidis2015public, bansal2016assessing, cunningham2019public, wilson2022non}. The broader literature on this topic consistently indicates that anticipated in-vehicle activities are not uniform but heavily influenced by contextual factors. For instance, Perterer et al. (2016) found that commuters clearly distinguish between journeys to work, where they typically engage in work-related tasks, and return trips home, characterized by more leisure-oriented activities~\cite{perterer2016activities}. Similarly, Pudāne et al. (2019) demonstrated that the variety of in-car behaviors in AVs significantly correlates with individuals' schedule constraints and daily routines~\cite{pudane2019will}. This contextual variability in NDRA preferences has been further substantiated by multiple survey-based studies~\cite{pfleging2016investigating, cunningham2019public, wadud2019fully}, highlighting the complex relationship between travel purposes and activity choices.

Despite these valuable insights into what people might do, how to reliably investigate future NDRAs faces significant methodological challenges. A primary issue is that most potential users lack practical experience with autonomous driving. This limits the predictive validity of conventional 'stated intention' approaches—such as user interviews, surveys, or observations of analogous transport modes~\cite{stevens2019using, pfleging2015driving}—as anticipated behaviors generated through speculation may differ markedly from actual future practices~\cite{sorensen1992thought, yang2020chicken}. Laboratory-based driving simulations attempt to bridge this experiential gap by offering more immersive environments~\cite{large2017design, hecht2020you}. However, while providing a degree of realism, these experiments often remain constrained by conventional vehicle designs and limited session durations~\cite{detjen2020wizard}, and crucially, they struggle to adequately capture the complex interplay between potential NDRAs, the specific travel context, and individuals' broader everyday activities and routines. Authentically embedding speculative future journeys within the rich, dynamic fabric of participants' unique daily lives remains a key difficulty for both simulation-based and conventional stated-intention methods.

In response to these challenges, particularly the need for deeper contextual understanding and experiential grounding beyond generalized scenarios or simulations, alternative methodological approaches have been explored. Participatory design workshops and co-creation methods, for instance, aim to involve users more directly in envisioning future scenarios~\cite{pfleging2016investigating, stevens2019using}, attempting to make speculation more concrete and user-centered by drawing more explicitly on user perspectives. Seeking insights from embodied experiences within a specific, prolonged context, Stampf and Colley (2024) utilized an autoethnographic approach during a 12-day recreational vehicle (RV) journey~\cite{stampf2024deriving}. Treating the RV trip as a surrogate for extended exposure, their situated analysis identified novel NDRAs and associated challenges particularly relevant to longer journeys and vehicles potentially used as living spaces, aspects difficult to capture otherwise. Further broadening the scope of interaction research, Stampf et al. (2024) also introduced a comprehensive design space for Cross-Traffic Interaction (CTI)~\cite{stampf2024move}, systematically organizing interaction possibilities between diverse traffic entities which can inform the design of NDRAs that leverage external connectivity or interaction with the wider traffic environment.

While these innovative approaches offer valuable steps towards more contextually rich or experientially grounded insights, limitations remain. Participatory workshops, though engaging, may still present somewhat generalized travel scenarios that don't fully embed the journey within the participant's specific, complex daily activity patterns. Autoethnographic surrogate studies, while providing rich, deep insights from the researchers' perspective derived from embodiment in a specific, extended setting, are inherently limited by that viewpoint and context. Similarly, structural tools like design spaces are crucial for mapping potential interactions but do not inherently bridge the gap to understanding how diverse individuals will personally adopt and experience these potential future activities within their unique daily lives. Crucially, there remains a gap for methodologies that systematically enable diverse participants to bridge their own unique, everyday lived experiences with speculative future AV scenarios through participatory and embodied exploration.

\subsection{Experience-Centered Design for Envisioning Future AV Experiences}
In HCI, Experience-Centered Design provides a valuable framework for understanding the intricate relationships between humans and technology~\cite{10.1145/3290605.3300634,10.1145/1957656.1957708,10.1145/3359594}. ECD conceptualizes experience as a dynamic, situated, and holistic process involving sensual, emotional, compositional, and spatio-temporal threads woven into everyday life~\cite{ECD}. This approach emphasizes experience not merely as an interaction outcome but as a complex phenomenon connecting past, present, and future, focusing on how interactions are felt, made sense of, and given meaning within an individual's broader life context~\cite{ECD}.

To explore experience in this way, ECD often employs participatory and performative methods, such as speculative enactments~\cite{10.1145/3025453.3025503}, role-playing~\cite{inproceedings123}, scenarios~\cite{davidoff2007rapidly}, and other participatory techniques~\cite{article123}. These methods aim not simply to collect data, but to create spaces for dialogue, reflection, and collaborative imagination, enabling participants to engage with potential futures grounded in their own values and perspectives. By eliciting narratives about habits, dispositions, and intuitions, ECD seeks to trace the 'continuity between past, present and future'~\cite{ECD}, making it particularly relevant for future-oriented inquiries where understanding expectations, desires, and concerns shaped by existing experiences is crucial.

Investigating potential NDRAs in future AVs involves aspects where ECD principles are applicable. First, such research requires exploring future possibilities; ECD's future-orientation and use of participatory, often narrative-based methods are suited for such exploration~\cite{ECD, 10.1145/3025453.3025503}. Second, understanding NDRAs involves considering their integration into complex daily routines and life contexts; ECD's core emphasis on situatedness and meaning-making within the broader fabric of life~\cite{ECD} offers a pertinent perspective for this contextual exploration. Third, the shift to AVs necessitates understanding the subjective, felt quality of potential in-vehicle life; ECD's focus on the holistic, multi-faceted nature of experience allows for investigating these qualitative dimensions~\cite{ECD}. Concepts such as Shedroff's (2001) view that future experiences build upon past ones~\cite{inbook232}, and Urry's (2007) perspective that travel is deeply embedded in a web of activities and relationships~\cite{urry2007mobilities}, further support applying ECD to leverage understanding of current travel practices as a foundation for exploring potential NDRAs.

While ECD principles appear well-aligned with the challenges of NDRA research, their limited application in envisioning NDRAs in AVs presents an opportunity to gain richer insights, particularly given the profound experiential shift autonomous driving entails. Therefore, this research adopts an ECD approach, specifically utilizing its focus on lived experience and context to investigate the intricate connections between participants' current daily routines, travel habits, and their envisioned NDRA engagement in future AV journeys, aiming to generate contextually-grounded insights for future design.

\section{Methodology} 
This study investigates potential NDRAs within future AVs by adopting an Activity-Based Travel perspective. This lens acknowledges that travel in AVs is unlikely to be an isolated segment but rather deeply intertwined with individuals' daily activity patterns and routines~\cite{axhausen1992activity, lyons2005travel}. Capturing this intricate relationship requires moving beyond traditional methods that often decontextualize travel. We employed ECD~\cite{ECD} for this research. ECD suits our future-oriented inquiry because it emphasizes understanding phenomena through lived experience, recognizing that expectations and desires for the future are grounded in past and present realities~\cite{mccarthy2004technology}. This aligns with our goal to explore how current travel practices and broader life contexts might shape future NDRA engagement in AVs. The study unfolded over three distinct phases (participant selection, sensitization, and participatory workshop) designed to progressively build understanding, moving from current experiences toward envisioned future scenarios.

\begin{figure}[htbp]%
  \includegraphics[width=\textwidth]{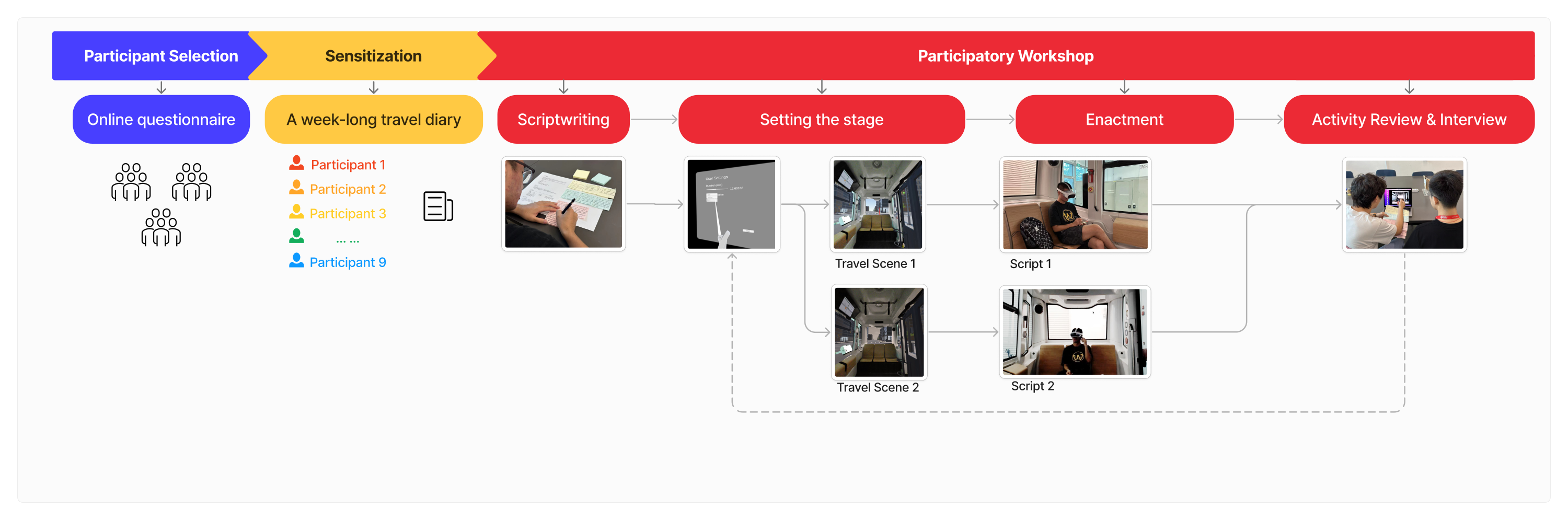}
  \caption{Representation of workshop design process}
  \label{fig:process}
\end{figure}

\subsection{Phase 1: Participant Recruitment}
Participant recruitment sought individuals with varied daily travel patterns and reflective capabilities. An initial online interest form and questionnaire were distributed through multiple channels including local community platforms, organizational networks, and social media. We utilized purposive sampling combined with snowball sampling techniques, encouraging initial respondents to share recruitment materials with potentially suitable contacts. This approach aimed to broaden the participant pool beyond the initial respondents. The questionnaire collected demographics, detailed travel routines (frequency, duration, modes, purpose), current travel activities, and initial thoughts on AV NDRAs. Following screening of the complete responses (resulting in N=67 initial respondents, 53 complete responses), nine participants were selected for the main study. The selection process focused on achieving representation across key characteristics relevant to travel behavior (e.g., age, occupation, travel modes, journey purposes) rather than statistical diversity, ensuring a spread of relevant perspectives for qualitative inquiry.

\begin{table}[!ht]
  \setlength{\tabcolsep}{4pt}
  \renewcommand{\arraystretch}{1.15}
  \caption{Participant characteristics and travel profiles.}
  \label{tab:Participants}
  \begin{tabular}{@{}cp{3.2cm}p{4.4cm}p{6.0cm}@{}}
    \toprule
    ID & Participant profile & Travel pattern & Technology, needs, and residential context \\
    \midrule
    P1 & 26; F; PhD student; scholarship & 6--7 trips/week; 30 min; public transit and bike-sharing & High technology familiarity; late-night commutes; university district \\
    P2 & 38; F; corporate employee; stable income & 5--6 trips/week; 30--45 min; private car and ride-hailing & Medium technology familiarity; diverse destinations; suburban area \\
    P3 & 45; M; entrepreneur; high income & 3--4 trips/week; more than 60 min; private car and rail & High technology familiarity; business trips; city center \\
    P4 & 22; M; student athlete; family supported & 3--4 trips/week; more than 60 min; public transit and chartered transport & Medium technology familiarity; sports equipment; university dormitory \\
    P5 & 32; M; technology manager; above-average income & 5--7 trips/week; 15--20 min; private car and bike-sharing & Very high technology familiarity; efficiency-focused travel; technology park vicinity \\
    P6 & 70; M; retired; limited pension & 2--3 trips/week; 60 min; public transit and walking & Low technology familiarity; medical appointments; suburban area \\
    P7 & 76; M; retired (part-time); supplemented income & 5--6 trips/week; 20--30 min; public transit and motorcycle & Basic technology familiarity; fixed work routes; urban-rural fringe \\
    P8 & 26; M; international researcher; stable income & 6--7 trips/week; 40 min; public transit & High technology familiarity; foreign environment; city center \\
    P9 & 25; F; student; research stipend & 1--2 trips/week; more than 60 min; ride-hailing & High technology familiarity; mobility impairment; city center \\
    \bottomrule
  \end{tabular}
\end{table}

\subsection{Phase 2: Sensitization}
The primary goal of this phase was to encourage participants to reflect deeply on their current travel experiences and surrounding contextual factors, thereby heightening awareness beyond habitual recall. Following contextmapping principles~\cite{visser2005contextmapping}, this phase prepared participants for subsequent generative activities by activating their tacit knowledge regarding travel. To facilitate this process, we developed a bespoke "Journey Canvas" tool, designed as a structured diary package (See Figure 2). Drawing inspiration from cultural probes~\cite{gaver1999design} and generative toolkits~\cite{sanders2000generative}, the Journey Canvas captured the holistic nature of travel experiences. Unlike standard travel logs that often focus solely on the trip itself, our canvas featured three dedicated sections to encourage broader reflection:
\begin{enumerate}
\item A Pre-Journey section captured activities, planning, considerations, and feelings before travel commenced.
\item An En-Route section documented the journey itself, including any NDRAs performed, travel conditions, interactions, and feelings during transit.
\item A Post-Journey section recorded activities, transitions, and reflections immediately after the journey concluded, explicitly linking the travel segment back to the participant's ongoing daily schedule.
\end{enumerate}

\begin{figure}[htbp]%
  \includegraphics[width=\textwidth]{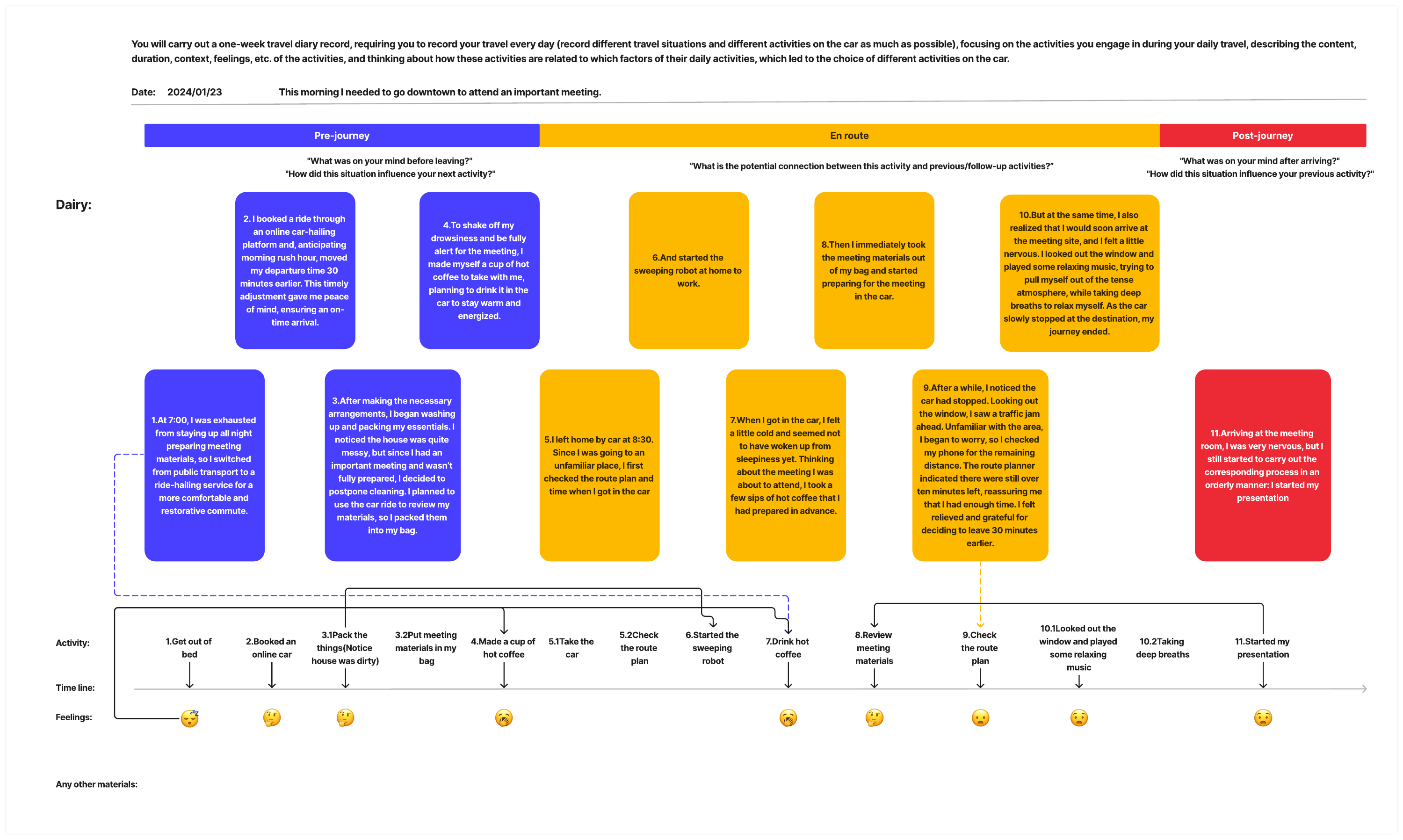}
  \caption{Journey Canvas for Sensitization}
  \label{fig:sensation}
\end{figure}

Several key design features supported rich data capture. A timeline structure facilitated detailed recording of activity sequences and durations, inspired by timeline mapping techniques~\cite{kujala2011ux}. Contextual prompts—open-ended questions embedded within each section (e.g., "What was on your mind before leaving?", "How did this journey affect your next task?")—stimulated reflection on the connections between travel and broader activities, addressing limitations of purely descriptive diaries~\cite{ettema2007multitasking}. Emotional fluctuations could be visually represented using customizable emoji templates inspired by models of affect~\cite{russell1980circumplex}. Furthermore, we encouraged multimodal expression through colored pens, stickers, and blank space for sketching~\cite{sanders2000generative}, alongside an Activity-Travel Mapping Area where participants could visually connect their activities and travel choices, reflecting an activity-based perspective~\cite{mokhtarian2001derived}.

Participants engaged with the Journey Canvas over a one-week period, documenting their daily travel experiences to capture a range of typical situations. For each journey, they completed all three sections using the tool's features to detail activity sequences, durations, locations, company, feelings, and relevant contextual factors (e.g., trip purpose, traffic conditions, connections to other activities), utilizing their preferred modes of expression (text, sketch, etc.). Researchers maintained brief email contact midway through the week to address queries and provide encouragement, ensuring participant engagement and data quality, consistent with diary study practices~\cite{bolger2003diary}. Crucially, the Journey Canvas served primarily as a sensitization and reflective tool to deepen participants' awareness for the co-creation phase, rather than constituting primary data for formal analysis within this paper's scope. The insights participants gained directly informed their subsequent scriptwriting in the participatory workshop.

\subsection{Phase 3: Participatory Workshop}
Following sensitization, the participatory workshop aimed to bridge participants' reflections on current experiences with the envisioning of future AV possibilities. This phase focused on translating situated knowledge into concrete future scenarios and exploring potential NDRA engagement through embodied enactment.
The workshop utilized several key materials:

\begin{enumerate}
\item A modified version of the Journey Canvas served as a Scriptwriting Tool. By removing the "En-Route" section and retaining only the "Pre-Journey" and "Post-Journey" components, this tool prompted participants to construct rich contextual backgrounds for future AV journeys, effectively creating a "stage" or "dramatic void"~\cite{bodker1999scenarios} for the travel segment to be explored via enactment; visual consistency with the original canvas leveraged familiarity.

\begin{figure}[htbp]%
  \includegraphics[width=\textwidth]{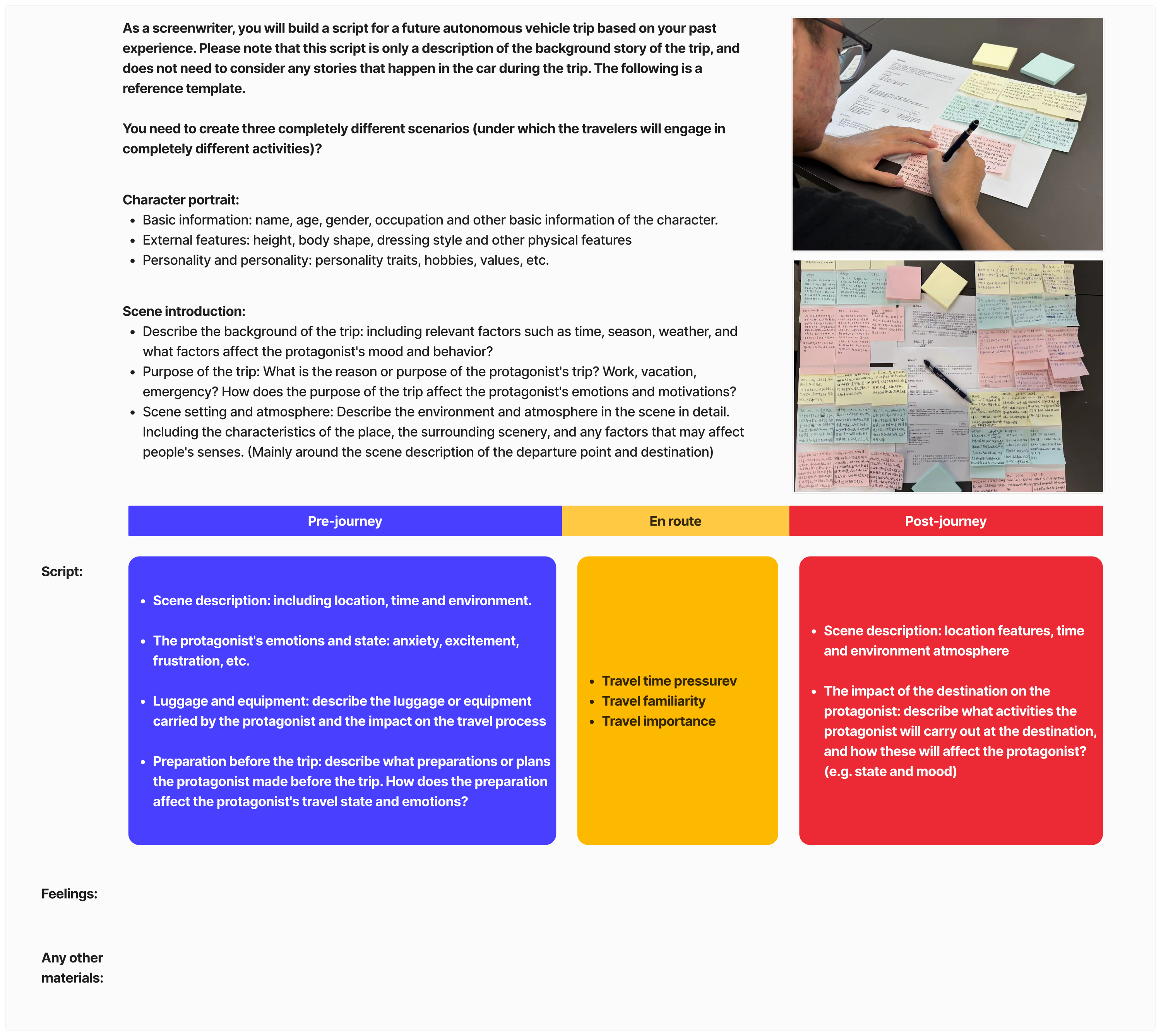}
  \caption{Journey Canvas for Scriptwriting}
  \label{fig:script}
\end{figure}

\item The central component was a MR Enactment Stage, comprising a physical mock-up approximating an AV cabin interior augmented with projected virtual travel scenes (e.g., city streets, highways) on surrounding screens. These virtual scenes were dynamically adjusted in real-time (e.g., time of day, weather, traffic density) based on parameters from each participant's script to enhance immersion and realism~\cite{brandt2000evoking}.

\item White Cards were provided during enactment as artifacts~\cite{mattelmaki2014happened}; participants used these blank cards to represent and physically place desired devices, interfaces, or functionalities that would support their envisioned NDRAs within the cabin space.
\end{enumerate}

The workshop procedure involved three main steps:
First, in the Scriptwriting stage, each participant authored three distinct future AV travel scripts within the workshop setting. They based these scenarios loosely on their documented past experiences but projected them into a future AV context, detailing the pre-journey situation, key journey parameters (purpose, destination, time constraints), and the post-journey context using the modified Journey Canvas structure, aligning with methods for creating rich, contextualized scenarios~\cite{rosson2002usability}.

\begin{figure}[htbp]%
  \includegraphics[width=\textwidth]{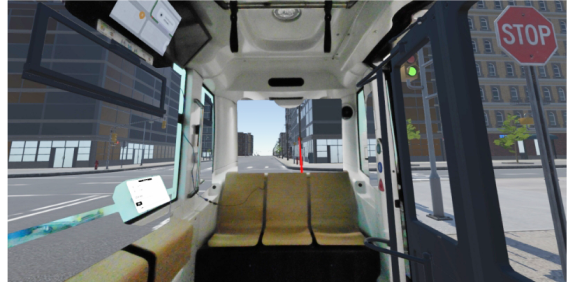}
  \caption{A morning with clear, open streets and sunny weather}
  \label{fig:street}
\end{figure}

Second, during Scenario Selection \& Enactment, each participant chose two scripts they found most compelling for enactment. The researchers configured the MR stage according to the selected script's parameters. The participant then entered the cabin mock-up and role-played their journey, performing anticipated NDRAs and interacting with the imagined AV environment as scripted, using the white cards to articulate specific needs by noting the function/device and placing the card accordingly. Researchers observed and video-recorded these enactments unobtrusively, employing embodied exploration principles~\cite{oulasvirta2003understanding}.

\begin{figure}[htbp]%
  \includegraphics[width=\textwidth]{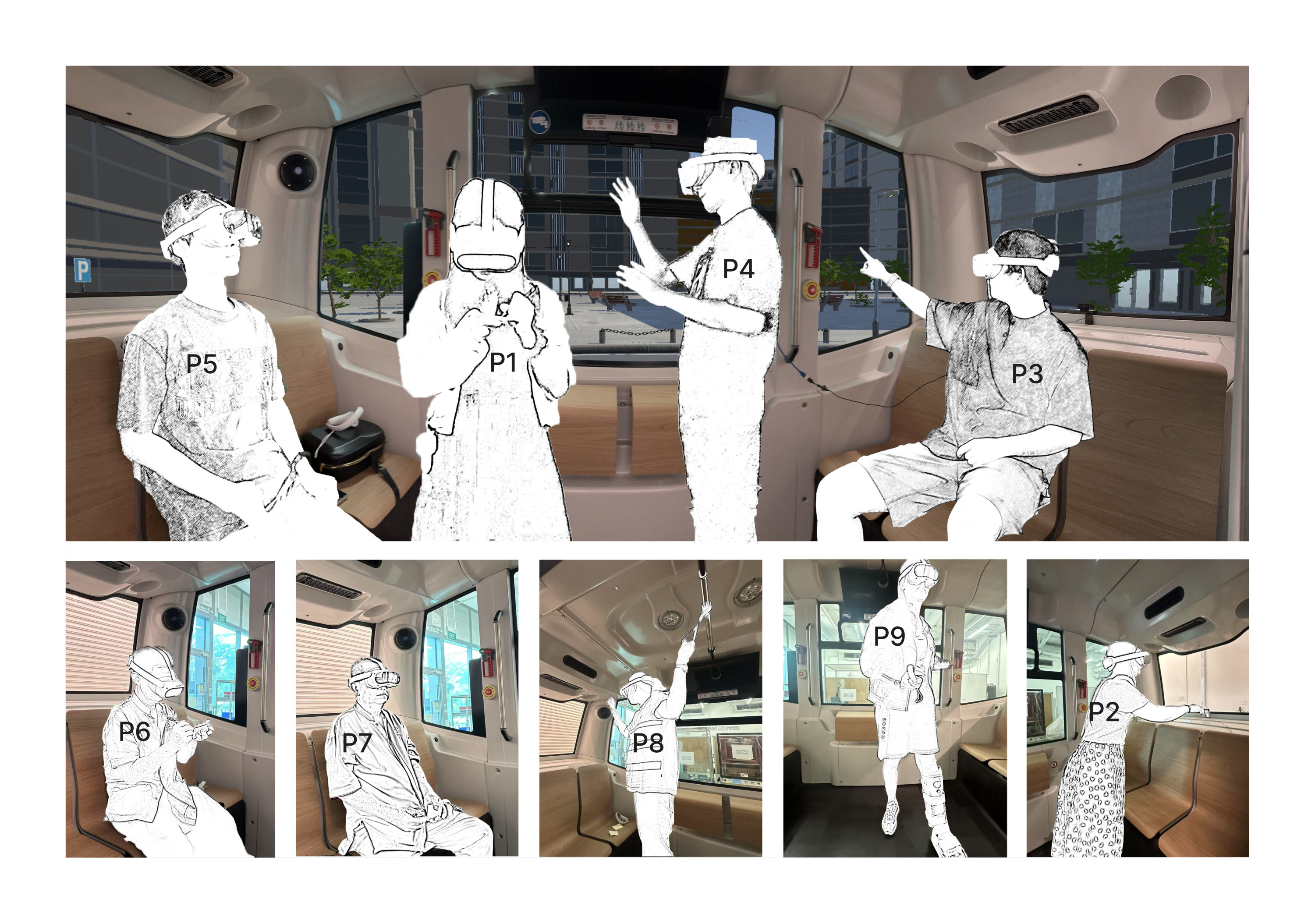}
  \caption{Mixed Reality Enactment within the Autonomous Vehicle (AV) Cabin}
  \label{fig:banner}
\end{figure}

Third, Post-Enactment Reflection occurred immediately after each enactment. A semi-structured interview, guided by video-stimulated recall~\cite{lyle2003stimulated}, facilitated discussion between the researcher and participant while reviewing enactment segments. This prompted reflection on the participant's actions, feelings, motivations, the perceived usability of envisioned NDRAs, and the rationale behind their white card placements. This enactment and reflection cycle was repeated for the participant's second chosen script.

\subsection{Data Analysis}
The analysis focused on the rich qualitative data generated during the co-creation phase, comprising 27 participant-authored scripts, video recordings and researcher field notes from the 18 enactment sessions (two per participant), 56 artifactual 'white cards' created during enactments, and transcripts from the 18 corresponding post-enactment semi-structured interviews (one conducted immediately following each enactment). We employed thematic analysis~\cite{braun2006using}, adapted to capture the processual nature of the study and the evolution of participants' perspectives as they moved from scripting to enactment and reflection. The analysis involved three main stages:

\begin{enumerate}
\item Analysis of Scripting Data: Initial analysis focused on the 27 participant-authored scripts. Open coding identified recurring themes related to how participants constructed future AV travel contexts, anticipated challenges, and envisioned potential NDRAs before the enactment phase. Two researchers coded independently, followed by discussions to compare codes and refine the initial thematic structure.
\item Analysis of Enactment Data: Video recordings of the 18 enactment sessions were analyzed, focusing on participants' behaviors, interactions with the space, expressed emotions, and particularly the use and placement of the 56 "white cards." We examined how NDRAs were performed, where supporting features were desired (indicated by card placement), and any moments of improvisation, hesitation, or problem-solving. This involved detailed observation logs and discussion between researchers to interpret the embodied actions in relation to the scripts and environment.
\item Integrated Analysis and Temporal Mapping: The final phase involved analyzing the 18 post-enactment interview transcripts using axial coding. We sought to connect themes emerging from the interviews with findings from the previous two stages. A key technique involved creating timeline-based visualizations for each participant, mapping their journey from initial reflections (Journey Canvas), through scripted scenarios, embodied enactment (video data, card placements), to final reflections (interviews). This temporal mapping helped identify convergences, divergences, and the evolution of participants' ideas about NDRAs and contextual needs throughout the study, paying attention to nuances and tensions expressed.
The entire analysis was iterative, involving regular meetings among the research team to discuss emerging themes, resolve discrepancies in coding, and ensure rigor and consistency in interpretation across the different data types and phases.
\end{enumerate}

\section{Findings} 

\begin{figure}[htbp]%
  \includegraphics[width=\textwidth]{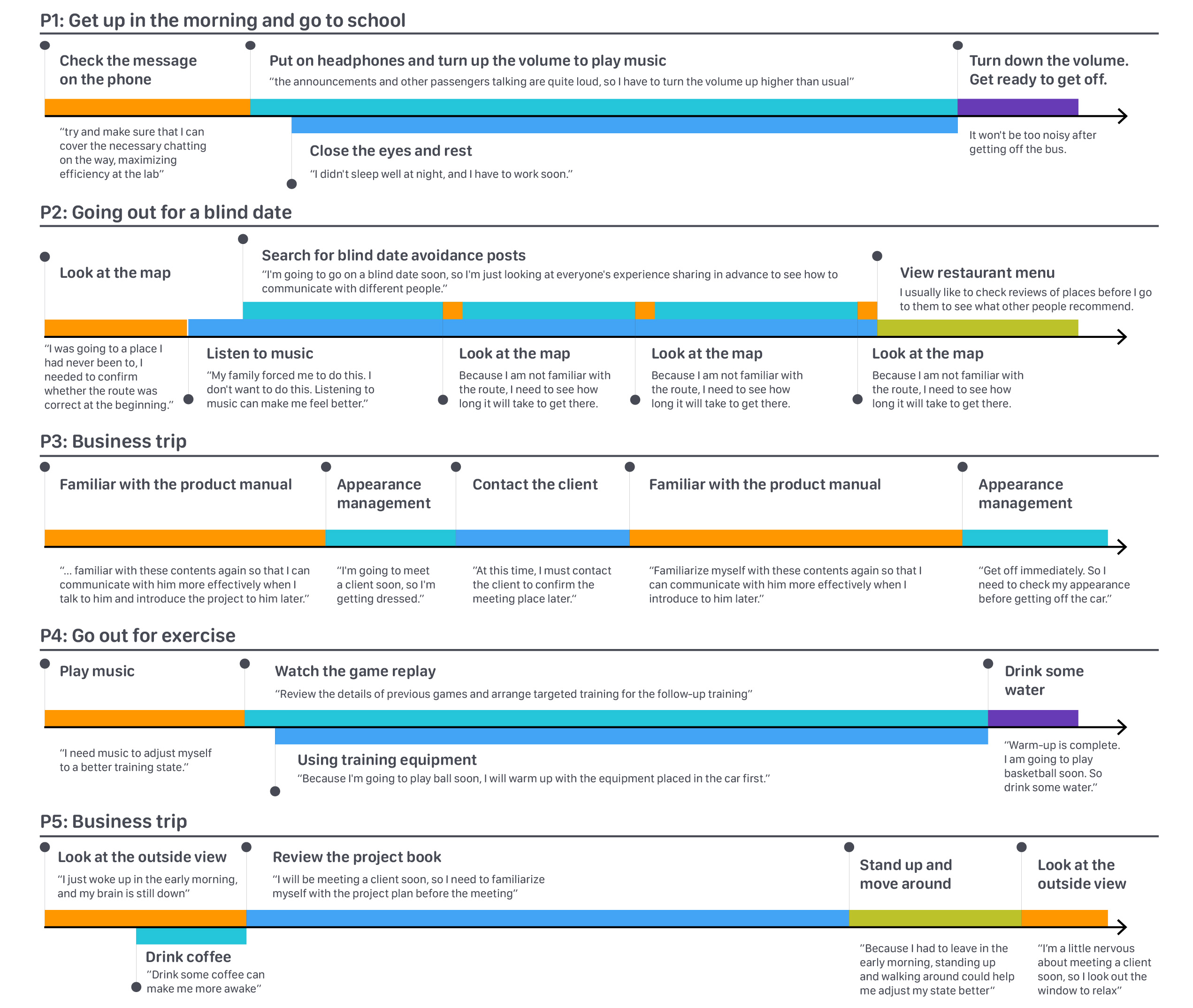}
  \caption{Records of In-Vehicle Activities Conducted by Some Participants While Traveling in Autonomous Vehicles}
  \label{fig:NDRA1}
\end{figure}

\subsection{Envisioning NDRA and FAV Scenarios with ECD}
\textit{In Enactments} \\
The enactment is regarded as the implicit expression of personal experience through immersing participants in manifold predefined roles and scenarios. The actions performed by the participants underscored what were considered potential factors and how these factors related to NDRAs. P4, for instance, enacted a scenario involving a midnight ride in an FAV for a business appointment. As the story unfolded, rich contextual routines naturally emerged, encompassing activities such as "confirming contracts over the phone" and "checking news on a smartphone". P8, enacting a photographer, spontaneously used his camera, an activity enabled by freedom from driving and the scenario's novelty. Some experiential nuances, which were not explicitly written in the script but emerged in the enactments, were also captured and interpreted as key manifestations of certain factors. In the same instance, P4 associated his long stare out the window with a concern about the "late time" which prompted him to adjust himself to "preventing a vacant state of mind". Serving as a captivating form of direct engagement, the enactments fostered an immersive and improvisational connection between participants and a future AV scenario. P9's enactment, drawing on personal experience with mobility impairment, vividly expressed specific accessibility needs and anxieties. Even physiological or emotional responses, such as yawning or vigilant reactions to in-car cameras, were spontaneously and naturally triggered. Conversely, P7's initial enactment reflected a more reserved engagement, perhaps influenced by his perception of external constraints like governmental regulations, demonstrating how pre-existing attitudes could shape performative interactions, which provided both participants and researchers with subtle and intriguing elements of the experiential aspects within the vehicle.\\
\textit{Through Inquiry} \\
Researchers' inquiries into the enactments opened up a series of dialogues around the user experience of being settled in an FAV, which encompassed participants' self-interpretations of actions in enactment, exploration of deficiency in experience, potential coping, and the consequences of solution implementations. A notable instance is evident in the interview with P1. When discussing her apparent idleness during the enactment, she explained that motion sickness had consistently restricted her activities on board. The researchers thus inquired about her expectations for responsive in-car services, which led to numerous intriguing hypotheses being raised, such as "ambient sounds like white noise," "adjustable seats with massage functions," "sunshades on windows," and "journey-related voice prompts." P7's responses to inquiry consistently framed the FAV within the context of Singapore's transport policies and high car ownership costs, revealing how socio-economic realities shaped his expectations primarily towards a public transport model. As such, the dialogue formed a responsive and continuously evolving transformative design resource. Meanwhile, due to the contextual presence, the expressed needs within the inquiry were multidimensional and deeply introspective.\\
\textit{Through Self-disclosure} \\
Participants' self-disclosure was typically manifested in a gradual progression throughout the post-study interviews, from reasoning about their actions to their feelings, desires, and values. This expands the topics involved, giving rise to vivid narratives connecting actual scenarios within future AVs. In the case of P2, for example, she drafted a scenario involving a reluctantly attended blind date, where she browsed through online posts on her phone in order to learn how to intentionally sabotage the blind date. She disclosed her inclination to scrutinize the daters and treat them conditionally during such encounters, which led her to propose creative in-car activities, such as "more likely to do makeup or change clothes in the car", and even "... seeing the date's appearance in my car and deciding how to dress based on that". Similarly, P9's disclosures evolved from practical concerns about mobility to revealing the emotional dimensions of her impairment, ultimately expressing how the lack of independent travel affected her sense of dignity and self-worth. She framed the potential of an accessible FAV not merely as a transportation solution, but as a means to restore her emotional well-being.\\
\textit{By Jumping out of Personal Experience} \\
As personal experiences serve as crucial foundations for envisioning the future, contemplating an alternative future with FAVs can be challenging for individuals accustomed to real-world scenarios. However, influenced by their experiences within the experimental FAV settings, participants transcended their immediate encounters to reflect on and reveal new understandings of possible NDRA futures. This shows a breakaway from the ritualized behaviors of traditional car travel to create a novel future mundane. Some participants indicated engaging in in-car behaviors during the enactment that they typically wouldn't do while commuting, such as humming a tune or standing up and moving around. They attributed these unconventional actions to a sense of "being in my space" afforded by the FAV. Building on this experiential shift, P8 transcended immediate impressions to consider longer-term behavioral adaptation, noting that "the initial novelty of the AV would eventually fade," necessitating the development of engaging activities or leading to repurposing commute time for tasks previously confined to home environments, such as handling emails. Some expressed a desire to place items which are not traditionally vehicle-mounted, such as a coffee machine or a liquor cabinet, within the private FAV. This notion stems from a reconsideration of the relationship between time utilization and in-car behaviors, suggesting the possibility of relocating activities traditionally conducted elsewhere to the interior of the vehicle. P3 articulated such transformative considerations regarding the positioning of the FAV in his daily affairs: "If I have to meet a client... I don't even need to arrange to meet at a coffee shop; I could conduct the entire conversation right here in this car." In perhaps the most radical departure from conventional vehicle perception, P7 transcended the very concept of transportation itself, reimagining the AV as serving entirely non-transit functions—conceptualizing it as an additional room for homes with adequate space (particularly noting this possibility in Malaysia's housing context) or as a temporary holiday cabin, completely detaching the vehicle from its primary mobility purpose and reconstructing it as a versatile living space.

\subsection{Artificial Scenarios: Endorsements and Critiques as Resources for Envisioning}
The artificial scenario created through the cabin and MR environment aimed to foster a connection between participants' past and future travel experiences. The endorsements and critiques regarding this "hardware" within the process were extensively articulated during interviews. Some participants commended the benefits of perception and cognition within the cabin and through the use of MR, highlighting it as an "immersive realism." P8 elaborated on this immersive quality, noting how the physical-virtual integration created "a really cool experience" that allowed him to interact with and respond to the environment as if it were a real AV, enabling a more concrete exploration of his future behaviors. Despite the cabin lacking any mobility functions, some participants even questioned researchers if the cabin was really on the move. P4 revealed the significance of these benefits in shaping anticipated meanings, which allowed him to contemplate his potential actions with "something tangible rather than just relying on my imagination".

In addition, some limitations surrounding the artificial scenario were also reported. Critiques emerged as the virtual and realistic contents, more or less, appeared to deviate from the participants' contextual situations or expectations, such as the "unrealistic appearance of some scenes" and "lack of unexpected situations" in MR.  P9 articulated a specific experiential limitation when attempting to use her personal device, noting that "looking at my phone in this environment makes me dizzy," which prompted her suggestion for integrated information displays within future AVs. The spatial dimensions also became a point of critique when P9 questioned whether "this feels a bit large for a car—is this realistic?" challenging the researchers' representation of a private AV's interior proportions. While these aspects did not align with their expectations, these critiques constructed a responsive space through dialogue, stimulating researchers' attention to new design opportunities. P4 asked whether the environment in MR could have an early morning setting following his script. This sparked discussions between him and the researchers on how the scenes should be portrayed and how he expected the FAV experience at that moment. Additionally, some participants mentioned discomfort regarding the inappropriate smells and uncomfortable seats within the cabin, prompting extensive expressions about coping for "customarily setting the interior smell in a specific scenario" and "having massage chairs or beds instead of seats." P7 specifically emphasized the sensory dimension of this customization, suggesting that "a massage chair would be perfect for long journeys" in his personally-owned AV, connecting physical comfort features to his envisioned private travel experiences.

\subsection{The Multifaceted Nature of NDRAs in Future FAVs}
Our analysis revealed three key themes concerning NDRAs in future FAVs: (1) the dynamic and interrelated nature of NDRAs, (2) the influence of pre- and post-journey contexts, and (3) the relationship between NDRAs and AV space. Insights from participants encompassing a wider range of ages, backgrounds, and physical conditions further enriched the understanding of this multifaceted nature. These findings collectively illustrate the complex, multifaceted nature of NDRAs in future FAVs. They highlight the need for a holistic, context-aware approach to understanding and designing these activities, considering not only the immediate journey but also the broader temporal and spatial contexts in which travel occurs. Moreover, they suggest a potential transformation in the very nature of travel, where the journey itself, facilitated by the unique affordances of AV spaces, becomes the primary focus rather than a means to an end.

\begin{figure}[htbp]%
  \includegraphics[width=\textwidth]{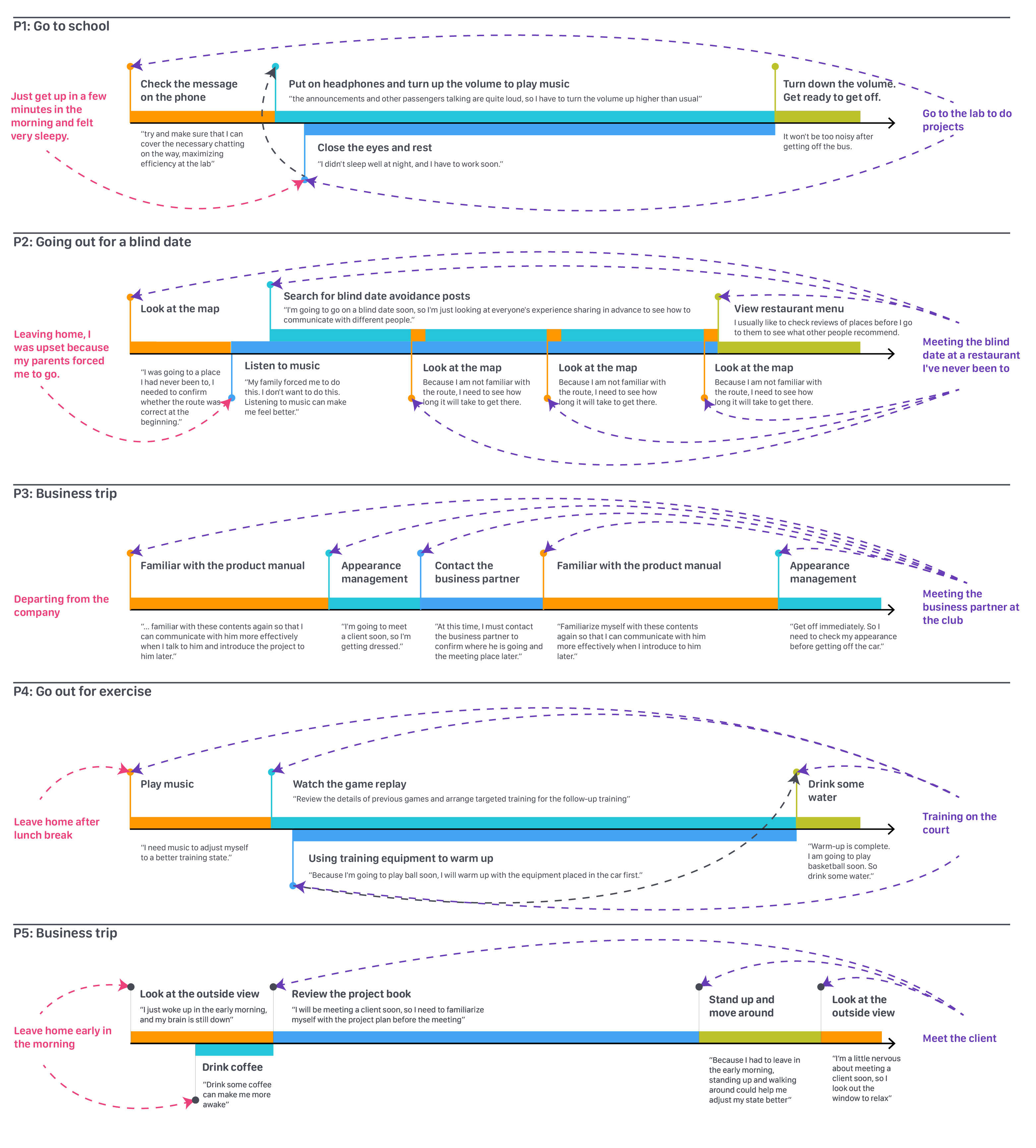}
  \caption{The Relationship Between In-Vehicle Activities Conducted by Some Participants and the Travel Context While Traveling in Autonomous Vehicles}
  \label{fig:NDRA2}
\end{figure}
 
\textit{NDRAs as a Dynamic Sequence of Interrelated Activities}\\
One salient observation was the existence of a primary NDRA during each journey, with other non-primary NDRAs woven intermittently to support or complement the primary NDRA. For instance, P1's use of headphones and music aimed to create a conducive environment for restful sleep, the primary NDRA during their morning commute. Similarly, P2's music listening appeared to alleviate anxiety while preparing for a blind date, and P5's initial gazing outside and coffee drinking seemed to facilitate mental alertness for reviewing project materials. P9 demonstrated this sequential approach by first initiating music playback to establish a calming atmosphere, then transitioning to active scenery observation, with both activities functioning as deliberate stages in her anxiety management sequence. This multitasking pattern suggests that non-primary NDRAs are not merely for passing time but rather serve as a means to optimize cognitive resources and craft an environment conducive to the primary NDRA. Notably, the findings challenge the traditional conceptualization of NDRAs as singular, isolated activities. Instead, NDRAs often manifest as a sequence or constellation of interrelated activities that evolve over the course of the journey. For example, P2's blind date preparation involved a progression from navigation tasks to information gathering and cognitive preparation, reflecting a strategic approach to managing uncertainty. Similarly, P4's exercise outing combined entertainment, analysis, physical preparation, and hydration, seamlessly integrating mental and physical readiness. This suggests that NDRAs are not merely isolated tasks but part of a holistic experience tailored to individual needs and preferences. Moreover, the data highlight the dynamic nature of NDRAs, with individuals adapting their activities based on evolving circumstances or needs. P5's incorporation of physical movement midway through the ride, potentially to combat lethargy, demonstrates this adaptability. P9's adaptive activity sequence included a deliberate pause in digital communication during continuous motion, strategically reserving message responses for stationary periods to avoid motion sickness, followed by resumption of different activities when moving again. Such flexibility in NDRA engagement suggests that individuals actively manage their experiences within the AV environment, optimizing their time and resources for personal goals.

\textit{"Prelude and Sequel to Travel" Matters in NDRAs}\\
Through interviews that explored the activities and the underlying reasons behind their enactments from three different scripts, the purpose of travel emerged as the most evident factor entwined with NDRAs. The participants' accounts suggested that different travel purposes play a significant mediating role in shaping diverse in-car activities and fostering distinct expectations of FAV design. For instance, P1 described how different travel purposes in her two scripts drove distinct NDRAs. When enacting a scenario of going to a bakery for a date, most of her in-car activities revolved around "finding nearby bakeries" and "staying in touch with her friends". In contrast, in the script where she commuted to the lab in the morning, she chose to reply to messages to ensure uninterrupted focus on her studies later on: "So, the idea is to try and make sure that I can cover the necessary chatting on the way, maximizing efficiency at the lab." Different participants exhibited similar NDRA and facility requirements for similar travel purposes. P3 and P5 both identified the need to review product manuals (primary NDRA) en route to business meetings to facilitate effective communication with business partners. Furthermore, specific preludes to travel were expressed to impact NDRAs. P1's lack of sleep the previous night led to closed-eye resting during her commute to school. Similarly, P5, facing an early morning business trip, opted for activities such as drinking coffee and looking out the window to wake up. P9's travel was significantly shaped by the prelude of sustaining a physical injury coupled with the sequel of an imminent hospital appointment, creating an anxiety-laden emotional trajectory that directly influenced her selection of calming NDRAs during transit. Moreover, participants like P2 (forced blind date - annoyance) and P4 (just woke up - fatigue) were emotionally influenced by prior events, leading them to play music in the car to adjust their mood. In subsequent interviews, P1 and P2 also indicated a habit of turning off their phones before sleeping, which resulted in them checking messages in the car before work in the morning. They viewed time in the FAV as a "buffer zone" between the travel prelude and subsequent activities. P8 articulated this buffer concept explicitly, describing how his travel sequence allowed for the strategic redistribution of activities between home and work sequels—specifically by transferring morning preparation tasks into the AV journey to extend his home sleep duration as a valued prelude benefit. While not all participants' activities were explicitly constructive, expressions like "I do this (task) because I want to do that (at a specific location)" were overtly or covertly interwoven in their explanations, indicating a goal-oriented approach to NDRA selection.
 
\textit{Reconfiguration of AV Space}\\
Our analysis reveals a complex relationship between NDRAs and the physical space of AVs, highlighting how these activities both shape and are shaped by the vehicle environment. The provision of resources (represented by our "white cards") allowed participants to envision novel AV configurations supporting a wide range of activities. This led to the conceptual migration of activities typically associated with other spaces into the AV environment. For instance, P4 reimagined pre-workout warm-ups, typically done at the gym, as in-vehicle activities supported by appropriate equipment. Similarly, P3 envisioned transferring personal grooming routines from the destination to the journey, enabled by the presence of a full-length mirror in the AV. P8 planned to migrate breakfast consumption and email handling into the vehicle, and P7 envisioned activities ranging from karaoke to using the AV as a temporary sleeping space. When AV spaces were not able to fully support desired activities, the participants demonstrated flexibility in adapting their NDRAs. For example, P2, unable to replicate her laboratory environment in the AV, opted for mental preparation and workflow planning instead. Even when offered the possibility of bringing lab equipment into the vehicle, she cited motion sickness as a barrier, illustrating how physical constraints of the AV space can influence NDRA choices. Participants frequently leveraged available spatial resources to create transitional or "pre-states" aligned with their destination activities. This included P1 napping to alleviate fatigue before studying, P1 and P2 using soothing in-vehicle services to transition to relaxation or sleep states when heading home, and P4 and P5 using music to motivate themselves en route to the gym. P9 employed music and scenery viewing as a transitional activity to mentally prepare for her hospital appointment, while P8 used the commute space as a transition from a busy home environment to either work or personal relaxation. The affordances of AV spaces prompted participants to reconsider their daily schedules and activity locations. This reimagining extended beyond mere adaptation to the AV environment, suggesting a potential paradigm shift where the mobility of AVs becomes secondary to their function as flexible activity spaces. In this context, NDRAs may no longer revolve around destination activities but become the core activities themselves, potentially inverting traditional travel demand. For example, P3 envisioned conducting business meetings within the AV due to the private, independent space it provides. P7’s concept of using the AV as a stationary room or holiday pod further illustrates this potential shift, where the vehicle’s value extends beyond mobility. This scenario illustrates how people might choose to travel specifically to engage in NDRAs within the AV space, rather than traveling to a destination for the activity. Such a shift could fundamentally alter our understanding of travel needs and purposes, with the AV becoming a destination in itself rather than merely a mode of transportation.

\subsection{Navigating Sociotechnical Boundaries: User Experiences with Future FAVs}
Our findings reveal that participants' understanding and expectations of FAVs extend beyond simple functional considerations, involving negotiations around complex sociotechnical boundaries. These boundaries manifest not only in relation to privacy and surveillance, safety and trust, accessibility and autonomy but are also deeply embedded within specific socio-cultural contexts. This negotiation process reflects how participants integrate technological possibilities with their personal experiences and values, thereby redefining the meaning of mobile space.

\textit{The Emergent Nature of FAV Space: Blurring Public/Private Boundaries and Contextual Constraints}\\
A central theme was the conceptualization of FAV space as possessing a unique intermediate nature, distinct from traditional public or private realms. Participants grappled with defining this new type of space, as articulated by P8 who described it as "neither private nor public, it's something in between," acknowledging both potential isolation from public view and awareness of corporate or state surveillance. This perception influenced expected behavior, with P8 noting a need to "maintain some kind of decorum."
Privacy expectations were consequently complex; participants like P8 expressed concerns about data usage while desiring user controls, seeking a balance rather than outright rejection of monitoring. P9 similarly positioned the FAV on a privacy continuum – potentially safer than ride-sharing but compromised by internal cameras, noting it was "not entirely a safe space for me."
This evolving spatial definition sometimes extended beyond mobility, with P7 envisioning the FAV as auxiliary living space, particularly feasible in contexts with ample space like Malaysia, suggesting you could "stay overnight... sleep in this room." However, such imaginings were consistently tempered by socio-cultural and environmental constraints. P7 readily acknowledged this use case was improbable in space-constrained Singapore, stating "Singapore really won't allow it." Similarly, P9 noted limitations based on social structures like family size. This highlights how the reimagining of FAV space, while expansive, remains grounded in and constrained by users' specific geographical, regulatory, and social realities.

\textit{Constructing Trust and Safety: Integrating Technical, Embodied, and Social Dimensions}\\
Participants constructed trust and perceived safety in FAVs through a multi-layered assessment involving technical, physical, social, and institutional factors. For some, like P6 who contrasted potential AV reliability with a negative experience with a taxi driver falling asleep, FAVs offered a potential increase in safety by mitigating human error. As P6 noted, "This autonomous driving is good, I won't worry."
However, trust extended beyond technical function to embodied experiences. P9 strongly emphasized the need for a smooth ride to avoid physical pain, illustrating how physical comfort and predictability are crucial components of perceived safety, especially for those with physical sensitivities. Her experiences also highlighted the limitations of purely technological safety measures and the unique value of human presence, particularly in perceived vulnerability scenarios. Recounting unsettling encounters on public transport, she concluded, "Cameras cannot compare to the sense of security provided by a person," suggesting that social presence co-creates safety in ways technology alone cannot replicate, particularly in specific cultural contexts like managing interactions with intoxicated individuals (her UK bus experience).
Finally, institutional trust played a significant role. P8 stressed the need for clear accountability and government oversight, stating the "government should be the entity owning and operating," linking trust in the technology to trust in the regulatory framework governing it. This indicates that user acceptance and perceived safety are contingent not only on the vehicle itself but also on the broader systems of governance and responsibility.

\textit{Social Implications: Balancing Empowerment, Accessibility, and Socio-Economic Realities}\\
FAVs were frequently envisioned as a technology with significant potential for empowerment and social inclusion, particularly concerning accessibility. P9 powerfully articulated how current mobility limitations impact not just physical movement but also emotional well-being, dignity ("I really dislike troubling others"), and social participation (missing events, limiting outings). For her, an accessible FAV (with features like ramps, handrails, a smooth ride) represented more than convenience; it signified restored autonomy and improved quality of life.
However, this potential for empowerment was often juxtaposed with the constraining realities of socio-economic factors. Participants P6 and P7 consistently highlighted the high cost of vehicle ownership in Singapore, leading them to conceptualize FAVs primarily within shared or public transport paradigms rather than as personally owned assets. As P7 noted, "Having your own car is quite taxing." This economic context directly influenced functional expectations, with P7 differentiating features feasible for private versus public FAVs, explaining that "Private ones can have massage chairs, public ones cannot."
While acknowledging the desirability of personal autonomy ("If you have your own car, you go where you want"), the perceived economic barriers significantly shaped their overall outlook and acceptance, sometimes fostering a sense of limited agency regarding future transport options. P7 expressed this sentiment clearly: "Whatever the government does, we just follow." This tension underscores how the potential social benefits of FAVs, such as enhanced accessibility, may be unevenly distributed or realized depending on prevailing economic conditions and ownership models.

\section{Discussion} 
Our research into NDRAs in FAVs reveals a complexity and interconnectedness that challenges traditional conceptions of in-vehicle activities. This section discusses the methodological implications of our ECD, followed by an exploration of how our findings suggest a potential paradigm shift brought about by AVs.

\subsection{Methodological Implications for AV Research}
\subsubsection{Scenario Materialization: Redefining Scenario Methods in Autonomous Vehicle Research}\ 

Scenarios have long been a powerful tool in HCI research for contextualizing and predicting future interactions~\cite{carroll2003making}. In AV research, the importance of scenario-based approaches is even more pronounced, given that mobility is inherently a highly contextualized activity. This section discusses how we extended the traditional concept of scenarios in HCI research by innovatively combining physical AV cabins with mixed reality (MR) technology to uncover the close relationship between people and their articulation about scenarios. We identify this relationship as scenario materialization. We explore how this materialization approach enhances the multidimensional representation of scenarios, particularly in revealing hidden factors, connecting the present and future, and contextualizing user needs.

The concept of materialization originates from design studies and has been widely applied in HCI research~\cite{koskinen2013design}. Dourish and Bell point out that materialization can transform abstract concepts into perceptible forms, thereby promoting deeper understanding and engagement~\cite{dourish2011divining}. In our research, materialization includes constructing physical AV cabins, creating MR overlays, and designing participatory travel scripts. This shift is not merely a methodological innovation but an exploration of future AV research paradigms. By transforming abstract scenario concepts into perceptible experiences, we can more deeply explore complex interactions between users and AVs. This approach resonates with Davidoff et al.'s~\cite{davidoff2007rapidly} findings on the importance of contextual factors in AV research while also expanding the boundaries of traditional scenario methods.

In the field of AV research, the application of scenarios has often been limited to narrative descriptions~\cite{hecht2020non,wilson2022non}. However, broader HCI research has adopted diverse approaches to constructing connections between humans, machines, and scenarios. For instance, Carroll's scenario-based design method emphasizes the importance of combining narrative, prototyping, and participatory design~\cite{carroll2003making}. We argue that it is necessary to introduce these richer scenario construction methods into AV research to more comprehensively capture the complexity of future AV experiences. Our approach creates multi-sensory, immersive scenario experiences by integrating physical space (AV cabin), virtual elements (MR overlays), and participatory narratives (travel scripts). This method not only includes space and characters but also reveals nuances of behavior, emotions, and social interactions in different scenarios. This "scenario materialization" represents a shift from purely narrative descriptions to multi-sensory experiences, offering new possibilities for AV research. Thus, we discuss the value and challenges of scenario materialization in three aspects:

\begin{itemize}
    \item [1.]
    Uncovering Hidden Factors: 
    Our immersive scenarios revealed factors that traditional research methods might overlook. For example, several participants instinctively grabbed their seats and leaned back when simulating vehicle acceleration, despite knowing they were stationary. This sparked discussions about in-vehicle safety cues. Such unconscious physical responses reveal potential user concerns that might not be mentioned in regular interviews. The value of this method lies in its ability to capture users' subconscious reactions and body language, which are difficult to obtain in traditional interviews or questionnaires. As Polanyi noted when discussing tacit knowledge: "We know more than we can tell"~\cite{polanyi2009tacit}. Our method makes this tacit knowledge visible and analyzable by creating a quasi-realistic environment. Spontaneous physical adjustments to manage pain during simulated motion revealed embodied knowledge about accessibility needs that might remain unexpressed in conventional interviews. Similarly, automatic reaching for seatbelts revealed ingrained safety behaviors that participants might not consciously articulate when merely discussing AV concepts. However, we must avoid viewing all bodily behaviors as meaningful interactions. For instance, we observed participants frequently looking around. Initial assumptions linked this to travel environment observation, but subsequent interviews revealed it was primarily due to curiosity about the virtual world. This risk of potential misinterpretation indicates that researchers need to interpret bodily behaviors cautiously, avoiding projecting their presuppositions onto participants.
    \item [2.]
    Connecting Present and Future: 
    By combining future scenarios with participants' current travel experiences, we created a continuum allowing for a more nuanced exploration of how current behaviors might evolve. This approach leverages the concept of experiential continuity, highlighting the critical role of cumulative and continuous experiences in learning and development~\cite{dewey1986experience}. In our study, participants imagined the future starting from familiar concepts, generating more realistic and feasible ideas for NDRAs. For example, when exploring business travel scenarios, some participants actively considered how to enhance cabin space to manage personal image, going beyond their initial simple speculations about in-AV activities. Comparative reflections across different regulatory contexts (Singapore versus Malaysia) exemplify how people draw on existing experiential frameworks to evaluate future possibilities. Similarly, the transfer of negative experiences with human drivers to positive expectations of automation reliability illustrates how current experiences form the foundation for future projections. This approach provides a framework for understanding how human behavior gradually adapts to technological changes. It avoids the pitfall of viewing future AV experiences as completely separate from current experiences, instead seeing change as a gradual process. This gradual shift in thinking provides valuable insights into how users adapt to and redefine new technologies. It demonstrates how users start from familiar behavioral patterns and gradually explore the possibilities brought by new technologies. This observation is highly consistent with the notion of "genealogical futures"~\cite{lindley2017implications}, emphasizing the continuity between present and future. However, overemphasizing continuity might suppress radical innovative thinking, as AV technology may bring fundamental behavioral changes that are difficult to predict by extrapolating current behaviors~\cite{voros2017big}. In some cases, participants struggled to think outside existing paradigms; when asked how to utilize increased travel time, some initially could only think of extending existing activities, like "watching more movies" or "sleeping longer." Therefore, we need to be alert to potential disruptive changes. The conceptualization of the AV as a stationary room extension rather than a mobility tool represents such a radical departure from conventional thinking that emerged only after extended engagement with our materialized scenario.
    \item [3.]
    Contextualizing User Needs: 
    Our scenario method provided a rich context for understanding how user needs change in an AV environment. Taking the after-work homebound scenario as an example, this scenario not only triggered discussions about dinner preparation but also extended to topics such as in-vehicle mini-kitchen facilities and daily time management. This multi-layered discussion illustrates the complex web of considerations that AV designers must navigate, echoing the multifaceted nature of AV user experiences~\cite{paddeu2020passenger,buckley2018psychosocial}. More importantly, it reveals the interactions and potential conflicts between different needs. As with the idea of preparing a simple dinner in the vehicle, participants soon realized this might conflict with vehicle safety and cleanliness maintenance requirements. Participants began to consider how to safely store and use kitchen utensils in limited space, and how to deal with food odors and residues, issues rarely considered in traditional vehicle design. This revelation of complexity challenges traditional feature-based design thinking~\cite{ulrich2016product,pahl1996engineering}, instead encouraging us to adopt a more holistic approach to understanding and meeting users' diverse needs in private AV spaces. The value of this method lies in providing a framework for capturing and analyzing the contextual nature of user needs, especially in the environment of private AVs. It demonstrates how multiple seemingly contradictory needs of a single user (such as relaxation, work efficiency, and personal care) may intertwine in specific scenarios. This insight resonates with actor-network theory~\cite{latour2007reassembling}, emphasizing the importance of understanding complex interactions between people, technology, and the environment, even within private AV spaces. In this way, we can develop more dynamic, more contextually sensitive user models, which are crucial for designing AV systems that truly meet the diverse needs of individual users.
\end{itemize}

As such, the materialized scenario reveals the advantages of motivating non-experts' in-depth engagement in AV research, which overcomes some inherent limitations of traditional AV research methods such as questionnaires, focus groups, and laboratory experiments. By creating simulated future scenarios, we enabled participants to more comprehensively experience and reflect on potential AV use contexts, thereby inspiring deeper engagement and creative thinking. The effectiveness of this approach was evident in the innovative concepts spontaneously proposed by participants. For example, ideas like in-vehicle fitness equipment and mobile bars naturally emerged from participants' interactions with the environment. These ideas not only showcase participants' imagination but, more importantly, reflect how potential users might reimagine the function and meaning of AV spaces. This spontaneous idea generation process demonstrates that participants are no longer passive research subjects, but active contributors, which is crucial for shaping the future~\cite{sanders2008co}. However, we also observed that not all participants were equally engaged in these creative activities. Some participants found it difficult due to a lack of creative skills or experience.

Another unexpected outcome of our research was its potential to influence perceptions of AV technology. Several participants reported that the immersive experience changed their views on autonomous driving, fostering greater understanding and acceptance. For example, one participant stated after experiencing a simulated AV commute: "I was initially skeptical about autonomous driving, but this experience showed me how it could integrate into daily life." Another mentioned: "Personally 'experiencing' an AV made me realize the time and space freedom it could bring." These responses indicate that the scenario method can play a role not only in design but also in preparing users for technological transitions. Through 'scenario materialization,' we were able to help participants visualize the changes that AV technology might bring, thereby reducing fear and resistance to the unknown.

\subsubsection{Enactment as a Powerful Futuring Technique in AV Research}\ 

In the field of HCI research, particularly when exploring future scenarios, "futuring techniques" play a crucial role~\cite{hoffman2013theorizing}. This study explores the potential of enactment as an effective technique for researching and shaping future AV interactions, positioning it as a core component of an experience-centered methodology. This section analyzes how enactment enables participants to connect with potential AV scenarios on both physical and emotional levels, thereby bridging the gap between abstract concepts and concrete experiences. We discuss how this technique can reveal unexpected user needs, preferences, and socio-technical implications of AV adoption. Our findings highlight the effectiveness of enactment in enhancing participants' adaptability and reflective capabilities, especially when faced with unfamiliar future scenarios.

\textit{Bridging the Gap Between Abstract Concepts and Concrete Experiences}\\
Our research demonstrates that enactment can effectively connect abstract concepts of autonomous driving with users' concrete experiences. This finding is particularly evident in scenarios involving NDRAs. While previous NDRA research has provided valuable insights, these studies are often limited to abstract predictions or conceptual discussions~\cite{kyriakidis2015public,hecht2020non,schoettle2014survey}. This abstraction makes it difficult for participants to relate these future scenarios to their current life experiences. Our approach attempts to bridge this gap by providing an "embodied" experience.
In the initial phase, many participants struggled to imagine how to effectively utilize time in autonomous driving, reflecting their abstract understanding of future technology. However, the enactment of "blank spaces" in the "journey canvas" played a crucial role, gradually transforming these abstract concepts into concrete experiences. These "blank spaces" not only served as catalysts for creativity but also became platforms for participants to embody future experiences. For example, a participant's improvised morning "wake-up" process and the idea of personal transformation activities during commute times revealed subtle user needs that might have been overlooked in previous research. Similarly, when participants engaged in activities such as mobile exercise or virtual meetings in the simulated AV environment, they connected with these potential future scenarios both cognitively and physically. The effectiveness of our approach resonates with previous research highlighting how concrete experiences can provide valuable insights for HCI studies~\cite{odom2014placelessness}. However, our study further suggests that such concrete experiences can not only provide insights but also potentially become important tools for shaping future technologies. By transforming abstract concepts into experiential concrete scenarios, participants were able to explore and understand future AV use contexts more deeply, generating richer, more meaningful insights that could substantially impact the design and development of future AVs.

Enactment serves as a bridge, connecting the abstract and concrete through:
\begin{itemize}
    \item [1.]
    Spatial Concretization: Participants move in simulated vehicle spaces, making the abstract "in-vehicle space" concrete.
    \item [2.]
    Action Practice: By physically executing actions (e.g., exercise movements), participants transform abstract "in-vehicle activities" into concrete bodily experiences.
    \item [3.]
    Scenario Simulation: By simulating different scenarios (e.g., morning commute), participants can relate abstract "future daily life" to current life experiences.
\end{itemize}

The effectiveness of this approach is reflected in the shift in participants' attitudes. For example, participants who previously expressed skepticism about effectively using AV travel time gradually showed positive engagement during the enactment process. In the sports travel scenario, they began exploring various exercise possibilities, such as "I could do yoga here" or "This space is suitable for stretching exercises." This transition is not merely a process from doubt to acceptance, but more importantly, a transformation from abstract concepts to concrete experiences. Notably, this transition from abstract to concrete was not always smooth. Some participants initially showed hesitation or even skepticism, reflecting their uncertainty when facing an unknown future. However, as the enactment progressed, many participants gradually exhibited more positive engagement, spontaneously beginning warm-up activities in the sports travel scenario. This transition should not be simply interpreted as acceptance of future technology, but rather viewed as a complex adaptation process, including participants' reimagining and critical thinking about future scenarios.

\textit{Enhancing Adaptability and Reflective Capabilities}\\
When faced with unfamiliar future scenarios, enactment not only enhanced participants' adaptability but also improved their reflective capabilities. Participants not only imagined future possibilities but also critically considered the potential impacts of these possibilities. This deep engagement and reflection enabled us to capture richer, more nuanced user perspectives, providing valuable insights for the design and implementation of future autonomous vehicles.
For example, in a simulated "blind date" scenario, a participant exhibited an anxious response. This reaction was not merely a response to a specific scenario but sparked a discussion about designing more humane emotional systems in autonomous vehicles. Participants began to ponder: "If the vehicle could sense the passenger's emotional state, how should it respond?" This reflection goes beyond simple functional considerations, touching on ethical and privacy issues in human-machine interaction~\cite{brey2000disclosive,nissenbaum2004privacy,picard2002computers}.
Enactment of navigating with a mobility impairment similarly triggered deeper reflections on accessibility beyond mere physical accommodation, extending to considerations of dignity and emotional well-being. Another example is when participants simulated long-duration AV rides, they began to consider the potential impacts of this new mode of mobility on social relationships and urban planning. One participant raised the question: "If we can work in the car, does it mean we would choose to live further away?" This reflection demonstrates how participants connect personal experiences with broader social impacts.
This deep-level reflection echoes research in the philosophy of technology~\cite{verbeek2015cover}, emphasizing that technology not only changes our behavior but also shapes our values and social structures. Our research method provides a unique platform for exploring this complex human-technology relationship.

\subsubsection{Multidimensional Data: Unveiling the Complexity of Future AV Experiences}\ 

In this section, we discuss the value of collecting multidimensional data as part of our ECD approach to studying NDRAs in AVs. This approach corresponds with Pelzer and Versteeg's advocacy for presenting futures through multiple sensory systems and forms of meaning-making~\cite{pelzer2019imagination}, addressing the complex functional, emotional, social, and ethical dimensions involved in AV technology~\cite{kun2016shifting,ekman2017creating,detjen2021increase}. We explore how combining various data collection methods - from text scripts to enactments and artifact creation - provides a rich and nuanced understanding of potential user experiences in future AVs. This multifaceted approach enables us to uncover hidden aspects of user interactions and reveal the complex socio-technical implications of AV adoption.

\textit{Deepening Multi-faceted Understanding of User Experiences}\\
Our research design progressed from text script creation to dynamic visual material production, scenario enactment, artifact creation, and culminated in in-depth interviews. This progressive research process stems from the recognition that the complexity of user experience necessitates multifaceted approaches~\cite{law2009understanding}. More significantly, it enables us to probe deeply into potential needs, emotional responses, and behavioral patterns within user-AV interactions.
As research activities deepened, participants' responses became increasingly complex and nuanced, revealing many hidden dimensions of user experience. For instance, while initial script writing primarily focused on describing AV scenarios, subsequent scenario enactments revealed deeper emotional responses: tension displayed during simulated emergencies reflected potential concerns about AV safety, while relaxed attitudes when experiencing entertainment features suggested potential expectations for AV comfort.
Through observing participants' body language and behavior in simulated environments, we discovered many embodied interaction needs that traditional questionnaires struggle to capture, such as unconscious gestures made when "conversing with the AV," providing inspiration for considering complementary visual feedback in voice interface design. In-depth interviews further revealed potential contradictions in user experiences, such as participants expressing a desire for AVs to increase efficiency and reduce travel time while also showing a need for more private space and "alone time" in the vehicle. This contradiction reflects the subtle balance users seek between pursuing efficiency and enjoying the journey itself.
The artifact creation phase allowed participants to concretize abstract ideas, revealing hidden needs related to specific use scenarios, such as designing flexible interior layouts to transition between work, rest, and social scenarios. This multi-layered, progressive research approach not only helped us understand users' explicit needs but, more importantly, revealed potential needs and behavioral patterns that users themselves might not be aware of.
Moreover, this multi-faceted approach presented challenges in data integration and analysis. To address this, we employed innovative data visualization techniques, annotating participant data on a timeline. This visualization method allowed us to simultaneously observe and analyze data from different research stages, uncovering potential patterns and insights. For instance, we could visually discern how changes in a participant's body language during scenario enactments corresponded with ideas expressed in in-depth interviews, or how artifacts they created reflected needs mentioned in earlier scripts. This approach not only facilitated cross-method data comparison but also helped us identify trajectories of user experience evolution over time.

\textit{Revealing Contradictions and Internal Conflicts}\\
Our research method uncovered significant inconsistencies between participants' verbal expressions, behaviors, and creations, providing a unique perspective for understanding users' complex attitudes towards AV technology. For instance, some participants expressed complete trust in AV technology during interviews, but displayed obvious discomfort and excessive vigilance during scenario enactments. This contradiction not only reflects the complex adoption process described in Rogers'~\cite{rogers2014diffusion} diffusion of innovations theory but also reveals deeper internal conflicts users face when confronting emerging technologies.
These contradictions may stem from multiple factors. Firstly, they might reflect differences between users' cognitive and emotional responses. While at a rational level, participants may understand and acknowledge the potential advantages of AV technology, at an emotional and instinctive level, they may still feel uncertain and anxious. This cognitive-emotional inconsistency echoes the risk-as-feelings theory proposed by Loewenstein et al.\cite{loewenstein2001risk}, which emphasizes the crucial role of emotions in risk perception and decision-making.
Secondly, these contradictions may reflect the gap between users' expectations of AV technology and their realistic cognition. Participants may hold idealized expectations for AV technology, but when faced with specific use scenarios, they begin to realize potential challenges and risks. This expectation-reality gap is closely related to the concepts of perceived usefulness and perceived ease of use emphasized in the Technology Acceptance Model (TAM)~\cite{davis1989perceived}.
Our multidimensional approach, combining physical activities (scenario enactments) and creative tasks (artifact design), can simultaneously touch upon the three levels of user needs proposed by Norman~\cite{norman2007emotional} in his emotional design theory: visceral, behavioral, and reflective, thus providing a more comprehensive picture of user needs.

\subsection{Reframing NDRAs in Autonomous Vehicles}
Our research into NDRAs in FAVs reveals a complexity and interconnectedness that challenges traditional conceptions of in-vehicle activities. This section discusses how these findings suggest a potential paradigm shift in our understanding of urban mobility and human-city interactions.

\textit{Redefining In-Vehicle Experiences: The Complexity and Interconnectedness of NDRAs}\\
Contrary to approaches often treating Non-Driving Related Activities (NDRAs) as discrete tasks~\cite{pfleging2016investigating,schoettle2014survey}, our findings reveal their complex and interconnected nature within FAVs. Rather than isolated, singular activities, NDRAs manifested as dynamic sequences of interrelated actions evolving throughout the journey. We observed passengers engaging in strategic combinations, such as working (primary activity) while listening to music (supporting activity), purposefully integrating actions in ways that extend concepts like 'equipped time'~\cite{lyons2005travel} by demonstrating how travelers might actively curate their journey experience. For instance, a passenger using music to alleviate anxiety is not merely multitasking but strategically modulating their affective state in response to the travel situation itself. This complexity reflects the potential for seamless transitions between work, entertainment, and personal care within FAVs, suggesting future designs may need to support diverse, potentially overlapping activity modes through adaptable spatial configurations and intelligent environmental controls.

\textit{Reshaping the Value of Travel Time: Temporal Context and Urban Mobility}\\
Our findings emphasize the crucial role that pre- and post-journey contexts play in shaping NDRAs, prompting a deeper reflection on the nature of urban mobility. The concept of FAVs as "buffer zones" between different life contexts offers a new perspective on the role of travel time in daily life. This resonates with concepts like the "gift of travel time"~\cite{jain2008gift}, but our findings particularly highlight the active agency travelers employ in shaping this time rather than passively receiving it. This perspective suggests a new urban rhythm where travel time becomes an integral, potentially productive or restorative, part of life, rather than simply 'down time'. It raises a profound question: Should we view urban mobility as a continuous experience—where the quality and nature of the journey are as significant as the origin and destination— rather than merely a process of moving from point A to B? This reconceptualization could have far-reaching implications for urban planning and transportation policies. For instance, if travel time is viewed as valuable experiential time, city designs might need to reconsider transportation infrastructure layouts to optimize this experience. Furthermore, this evolving understanding of time could influence our perception of urban rhythms. FAVs might become mediators between personal schedules and city tempos, potentially enabling new time management strategies and lifestyles where travel time is strategically used for self-improvement, relaxation, or social connection.

\textit{Blurring Boundaries: FAVs as Dynamic Activity Environments for NDRAs}\\
The physical space of FAV plays a crucial role in shaping NDRA possibilities and user expectations, as our findings demonstrate. The potential for activity migration into FAV spaces suggests a blurring of boundaries between vehicles and other life domains. This extends beyond the concept of the "mobile office"~\cite{laurier2004doing} to encompass a broader range of activities and hints at a paradigm shift: the mobility function of FAVs may become secondary to their role as flexible activity spaces, reflecting evolving conceptualizations of travel time use~\cite{lyons2005travel}.
Our study revealed instances where participants fundamentally reconceptualized the autonomous vehicle itself. Beyond enhanced transportation, participants envisioned AVs as flexible, mobile extensions of other life domains—potentially transforming into temporary stationary living spaces, similar perhaps to emerging notions of mobile 'third spaces'~\cite{oldenburg1999great}. This dissolution of boundaries represents a profound shift in how we might understand vehicles and mobility in the future~\cite{urry2007mobilities}. These reimaginings challenge fundamental assumptions about what constitutes a "vehicle" and suggest entirely new design paradigms may be necessary to fully realize the potential of autonomous technology emerging from user experiences.
This perspective, derived from our participants' envisioning, challenges basic assumptions about the nature of travel and vehicle design. If FAVs are viewed as mobile extensions of urban space rather than mere transportation tools, this could lead to radical changes in urban dynamics and travel patterns, an area actively discussed in AV impact studies~\cite{milakis2017policy}. As our findings suggest, FAVs might become mobile meeting rooms, entertainment centers, or even temporary living spaces, fundamentally altering our understanding of "vehicles."
This potential shift could also influence urban planning and architectural design, as speculated elsewhere~\cite{stead2019automated}. If FAVs become extensions of urban space, we might need to rethink the function and design of buildings. Future office buildings, for instance, might serve more as docking stations and service centers for FAVs rather than traditional workplaces. This transformation could lead to more flexible and dynamic urban layouts, where fixed and mobile spaces complement each other, creating new urban experiences.

\subsection{Beyond Function: Discussing the Sociotechnical Significance of FAVs}
Our study, by exploring participants' lived experiences and future visions, moves beyond purely functional assessments of autonomous vehicles. The findings illuminate how the meaning and acceptance of FAVs are not inherent in the technology but are actively co-constructed through complex negotiation of sociotechnical boundaries. This negotiation process, deeply embedded in personal values and socio-cultural contexts, challenges techno-centric perspectives and highlights the need for more nuanced, human-centered understandings of autonomous mobility futures. We discuss two central themes emerging from this negotiation: the constitution of the FAV as a novel experiential and relational space, and the contested nature of FAV futures situated within broader societal dynamics.

\textit{Constituting the FAV Experience: Situated Meaning-Making in a Hybrid Space}\\
A key insight from our research is that the FAV is experienced not simply as a mode of transport, but as an emergent hybrid space where established norms of privacy, behavior, and safety are actively renegotiated. Participants' conceptualization of the FAV as "neither private nor public" (P8) signifies more than just a physical location; it represents a site of social meaning-making where users grapple with ambiguity~\cite{urry2007mobilities}. They draw upon existing spatial logics (home, public transport, private car) but find them insufficient, leading to the formulation of new, context-specific behavioral expectations ("maintain decorum") and complex privacy considerations that balance personal enclosure against perceived surveillance~\cite{nissenbaum2009privacy}. This active interpretation contrasts sharply with design approaches that assume stable, predefined user responses based purely on technical features.
Furthermore, our findings underscore that trust and perceived safety are relational and embodied achievements, not simply technical attributes. While acknowledging potential benefits of automation (P6), participants constructed safety through a holistic lens incorporating ride quality (P9's physical comfort), the perceived social presence (or absence) of humans (P9's critique of cameras vs. people), and crucially, trust in the overarching institutional framework (P8's emphasis on governance). This resonates with sociotechnical perspectives arguing that trust in complex systems is distributed across technology, human actors, and organizational structures~\cite{wynne1996may, zucker1986production}. It challenges safety paradigms focused predominantly on technical reliability~\cite{kalra2016driving} by demonstrating that users' felt sense of security is deeply intertwined with social context, embodiment, and governance – factors that must be central to user-centered AV design. The FAV, therefore, emerges not just as a technology to be 'accepted' but as a complex relational space requiring ongoing negotiation and sense-making by its occupants.

\textit{Contested Futures: Navigating Potential, Politics, and Place in FAV Deployment}\\
While participants envisioned expansive possibilities for FAVs—reimagining them as accessible tools for social inclusion (P9) or even extensions of living space (P7)—these technological imaginings were consistently interwoven with, and often constrained by, tangible socio-political and contextual realities. This highlights that FAV futures are not predetermined outcomes of technological progress but contested terrains where potential benefits are negotiated against existing social structures, economic limitations, and cultural norms.
The vision of FAVs enhancing accessibility and restoring dignity for individuals with mobility impairments (P9) powerfully illustrates the technology's emancipatory potential. This aligns with calls for leveraging mobility innovations to foster mobility justice~\cite{sheller2018mobility}. However, this potential was immediately complicated by participants' acute awareness of socio-economic barriers (P6, P7 on Singapore's car ownership costs). The resulting stratification of expectations (features available in private vs. public models) suggests that without deliberate policy intervention, FAV deployment risks mirroring or even exacerbating existing inequalities, echoing critical perspectives on technologically driven urban fragmentation~\cite{graham2002splintering}. This tension reveals that accessibility is not just a technical design challenge but a socio-political issue embedded within broader economic contexts. Therefore, fostering genuine empowerment in the FAV context might mean focusing on granting 'more contestability' to diverse users, ensuring potentially marginalized groups can actively participate in shaping how these mobility systems integrate into their lives~\cite{xue2024should}. Moreover, the study strongly indicates that FAV adoption and meaning will be highly context-dependent, challenging notions of technological universalism. Participants' contrasting visions based on geographical space (P7: Malaysia vs. Singapore), governance expectations (P7 vs. P8), and cultural norms demonstrate that FAVs will be 'domesticated' or co-produced differently across diverse settings~\cite{jasanoff2004states,silverstone1996design}. Existing transport cultures and regulatory environments create path dependencies influencing how FAVs are integrated and perceived. This finding urges a shift away from designing generic FAV solutions towards contextually sensitive approaches that acknowledge the profound influence of place, culture, and politics on how technologies are ultimately lived and experienced~\cite{suchman2011anthropological}. The future of autonomous mobility, therefore, appears less as a singular technological trajectory and more as a mosaic of locally negotiated sociotechnical arrangements, each carrying its own set of possibilities and political implications~\cite{winner2017artifacts}.

\subsection{Towards a New Paradigm of Human-City Interaction? Implications for Research and Design}

The complex ways participants negotiated the sociotechnical boundaries of FAVs—redefining space, constructing trust relationally, and balancing potential empowerment against contextual constraints—point towards the possibility of a fundamental paradigm shift in understanding mobility itself. Our findings suggest a move beyond viewing travel simply as efficient A-to-B transit, towards conceptualizing it as a rich, multi-layered lived experience deeply integrated with the fabric of daily life and the urban environment. In this emerging paradigm, FAVs cease to be mere vehicles and potentially become dynamic mediators in the human-city relationship.

This perspective necessitates a shift in research and design focus from optimizing purely functional aspects towards understanding and enhancing the quality of the holistic experience during mobility. It involves foregrounding the emotional, social, and embodied dimensions of travel, which became evident through participants' enactments and reflections. Such a focus resonates strongly with Third Wave HCI's emphasis on technology's role in the broader tapestry of human experience, beyond instrumental, task-oriented interactions~\cite{bodker2006second, harrison2007three}. Designing for FAVs, from this viewpoint, becomes less about managing a driving task replacement and more about crafting meaningful environments for living, working, and connecting while mobile, situating the vehicle within a broader human-vehicle-home integrated ecosystem~\cite{xue2025my}.

Navigating this potential future may require individuals to develop new sociotechnical competencies and mobility practices. Learning how to effectively blend activities, manage personal boundaries within semi-private mobile spaces, interact with intelligent vehicle systems, and leverage mobility for well-being could become crucial aspects of engaging with future urban environments. Understanding the emergence of these practices, alongside the new social norms and etiquettes they might entail, presents a critical area for future inquiry.

Consequently, this research opens up a vital interdisciplinary research agenda at the intersection of HCI, transportation studies, urban planning, and social sciences. Key directions include:

\begin{itemize}
    \item [1.]
    Investigating the design of adaptive in-vehicle experiences that cater to the dynamic sequences of activities and shifting contexts identified in our findings, moving beyond static interior concepts.
    \item [2.]
    Exploring FAVs as mobile activity platforms and their potential impact on work-life patterns, social interaction, urban accessibility, and the very definition of public and private space.
    \item [3.]
    Developing context-aware intelligent systems for FAVs that move beyond navigation to support occupants' activities, manage ambiance, and facilitate seamless transitions between mobility and other life domains, while addressing the complex trust and privacy concerns highlighted.
    \item [4.]
    Examining the co-evolution of FAV technology with urban form, policy, and social equity, ensuring that advancements contribute positively to mobility justice and human well-being rather than exacerbating existing inequalities.
\end{itemize}

In conclusion, while the precise contours of the future relationship between humans, autonomous vehicles, and cities remain emergent, our study strongly suggests it will be characterized by a deep interweaving of mobility, space, context, and lived experience. Embracing this complexity, moving beyond functionalism towards a holistic, human-centered, and contextually sensitive approach, is crucial. It calls for collaborative, interdisciplinary efforts to guide the development of autonomous mobility not just as a technological advancement, but as a means to foster richer, more equitable, and more meaningful urban lives.

\section{Limitations}
While this study offers valuable insights through its innovative methodology, several limitations should be acknowledged. First, the qualitative nature of the research, involving nine participants recruited via purposive sampling primarily from a university context, prioritizes depth of understanding over statistical generalizability. Although diverse in age and occupation, the sample may not fully represent the breadth of potential AV users across different socio-economic strata, cultural backgrounds, or geographical locations beyond the Singaporean context that implicitly framed some discussions. Second, the use of a stationary cabin mock-up augmented with MR, while designed for immersion (Scenario Materialization), inherently differs from the dynamic experience of actual vehicular motion. This artificiality, despite its benefits in facilitating enactment, might influence participant behavior and the translation of findings to real-world AV deployment. The motion sickness reported by some participants in the simulation highlights this experiential gap. Third, the study captures perspectives at a specific point in time. As AV technology and societal familiarity evolve, individuals' perceptions, expectations, and desired NDRAs are likely to change, necessitating longitudinal research to track these shifts. Finally, while our ECD approach aimed to mitigate speculative responses, the futuring nature of the inquiry means findings reflect envisioned possibilities grounded in current experiences, which may diverge from actual adopted practices once FAVs become commonplace. These limitations, however, also define avenues for future research aiming for broader representation, enhanced simulation fidelity, and longitudinal tracking of AV adoption.

\section{Conclusion}
This paper presented an ECD approach, employing sensitization through bespoke Journey Canvases, participant-authored scenarios grounded in lived experience, and embodied mixed-reality enactment, to investigate potential NDRAs within FAVs. Moving beyond conventional methods often limited by speculative responses or decontextualized settings, our methodology facilitated a deep, contextually-grounded exploration of how individuals might integrate FAV travel into the fabric of their daily lives.

Our findings challenge simplistic views of "travel time use." We demonstrate that NDRAs are typically not isolated tasks but dynamic sequences of interrelated activities. These sequences are strategically orchestrated by individuals and profoundly shaped by the "prelude and sequel" to travel—the broader temporal context of daily routines, schedules, emotional states, and goals—positioning the FAV journey as an integrated segment of life rather than a disconnected interlude. This understanding was often accompanied by participants actively reimagining the AV's physical space, migrating activities from other domains and leveraging the environment as a transitional or preparatory zone, hinting at vehicles becoming multifunctional spaces beyond mere transportation.

Furthermore, the research illuminated the negotiation of complex sociotechnical boundaries surrounding FAVs. Participants conceptualized the FAV as a hybrid space demanding new considerations of privacy and behavior, constructed trust through multifaceted assessments encompassing technical, embodied, and institutional dimensions, and weighed the technology's potential for empowerment (particularly accessibility) against tangible socio-economic and contextual constraints. This underscores that AV adoption and meaning are highly context-dependent and locally negotiated.

Collectively, these findings advance our understanding by reconceptualizing NDRAs as dynamic, context-integrated sequences and illuminating FAVs as complex sociotechnical mediators—perspectives effectively surfaced through our ECD methodology, particularly its use of reflective tools and embodied enactment.

Ultimately, our work advocates for a shift towards understanding and designing for automated mobility as a rich, lived experience. This demands human-centered and contextually sensitive approaches that acknowledge the intricate interplay between technology, individual lives, and societal structures. Such a perspective is crucial for guiding the development of FAVs not merely as technological advancements, but as means to foster more meaningful, equitable, and well-integrated urban futures.

\bibliographystyle{unsrt} 
\bibliography{references}

\end{document}